\documentclass[12pt,a4paper]{article}

\usepackage{jheppub}
\usepackage{epsf}
\def\half{\frac{1}{2}}
\def\sumint{\hbox{$\sum$}\hspace{-1.2em}\int}
\def\fsumint{\hbox{$\sum$}\hspace{-1.2em}\int'}
\def\sumintsmall{\hbox{\scriptsize$\sum$}\hspace{-0.8em}\int}
\usepackage{slashed}
\usepackage{graphicx,amsmath,amssymb}

\def\Eq#1{Eq.~(\ref{#1})}
\def\be{\begin{equation}}
\def\ee{\end{equation}}
\def\bea{\begin{eqnarray}}
\def\eea{\end{eqnarray}}
\def\OO{{\cal O}}

\def\ZT{Z_{_{T}}}
\def\Nzero{N_0}
\def\None{N_1}
\def\Nhalf{N_{^{\!\frac{1}{2}}}}

\def\insertion{\begin{picture}(6,6)\thicklines\put(0,0){\circle{8}}
               \put(-4.6,-3){${\bf\times}$}
               \end{picture}}
\def\nottp{p\hspace{-0.44em}/}
\def\nott#1{\setbox0=\hbox{$#1$}                
   \dimen0=\wd0                                 
   \setbox1=\hbox{/} \dimen1=\wd1               
   \ifdim\dimen0>\dimen1                        
      \rlap{\hbox to \dimen0{\hfil/\hfil}}      
      #1                                        
   \else                                        
      \rlap{\hbox to \dimen1{\hfil$#1$\hfil}}   
      /                                         
   \fi}                                         %

\title{Thermodynamical second-order hydrodynamic coefficients}

\author{Guy D.\ Moore and Kiyoumars A.\ Sohrabi}

\affiliation{Physics Department, McGill University, 3600
rue University, Montr\'eal, QC H3A 2T8, Canada}%

\date{July 2012}

\abstract{Transport coefficients in non-conformal second-order
hydrodynamics can be classified as either dynamical or
thermodynamical.  We derive Kubo formuale for the thermodynamical
coefficients and compute them at leading perturbative order in a theory
with general matter content.  We also discuss how to approach their
evaluation on the lattice.
}

\begin{document}

\maketitle

\section{Introduction}

The theory of QCD is weakly coupled at short distances or high
temperatures, but strongly coupled at long distances or low
temperatures.  One of the major goals of both the experimental and
theoretical programs in QCD has been to understand
how quickly this transition occurs and at what energy.
A major purpose of the heavy ion collision
program was to see if weak-coupling behavior emerges at available
energies.  Similarly, lattice studies have investigated how close
thermodynamic properties come to their weak-coupling values as a
function of the temperature.

Broadly speaking, we can divide properties of thermal QCD into two
categories:  dynamical and thermodynamical.  Most of our information on
dynamical properties is from experiment.  Experiments show \cite{experiments}
that at available temperatures, QCD displays excellent fluid behavior
with remarkably low viscosity \cite{ideal_hydro}.  This is
very different from weak-coupling behavior \cite{Baym,AMY6}, but roughly
consistent with strong-coupling behavior in similar theories which we
can solve \cite{Policastro,Kovtun}.  The story for thermodynamic
properties, where most of our information is from the lattice, is more
complex.  At $T\sim 150$ to 200 MeV, thermodynamic properties such as
pressure, baryon susceptibility, and $\langle \bar\psi \psi \rangle$
show strong temperature dependence and are far from their weak-coupling
values \cite{Lattice}.  As temperature rises most
thermodynamic quantities approach weak-coupling behavior, but at
different rates.  Quark number susceptibilities come close to
weak-coupling behavior already at a few $T_{c}$ \cite{Lattice,Fodor}.
Cross-correlations between strange and light quark numbers
transition change from the expected behavior in a hadron gas to the
behavior of a weakly coupled plasma over this same temperature
range.  The pressure takes rather longer to approach weak coupling
behavior \cite{KarschEjiri}.

We feel that the more dynamical and thermodynamical quantities we have
available, the more complete and nuanced a picture of the strong to weak
coupling transition we can obtain.  With this in mind, we advocate
investigating the so-called second order hydrodynamic coefficients and
their coupling dependence.  As we will argue, some of these coefficients
are thermodynamical and can be computed on the lattice.  They also have
simple weak-coupling behavior.  Indeed, the main goal of this paper will
be to compute their leading-order weak-coupling behavior in a general
theory.  We will also discuss what would be involved in evaluating them
on the lattice.

In the the next section we will review second-order hydrodynamics
and explain how some of the coefficients of this theory are
thermodynamical.  Since at least the mid-rapidity regions in
high-energy heavy ion collisions deal with QCD at small quark-number
chemical potentials, and since the lattice can only deal well with
the case where chemical potentials vanish, we will assume vanishing
quark-number chemical potentials.  We also ignore magnetic fields.
However, QCD is far from a conformal theory in the interesting
energy regime, so we will not assume conformal symmetry.  In this
case there are three independent second-order hydrodynamic
coefficients which are thermodynamical in nature
\cite{Bhattacharyya,Minwalla2}.  In the notation of Romatschke
\cite{Rom_entropy,BRSSS} these are $\kappa$, $\lambda_3$ and
$\lambda_{4}$.  We compute their values at weak coupling and
vanishing masses in Section \ref{sec:weakcoupling}; see particularly
\Eq{eq:kappa} and \Eq{eq:lambda3}.

As we remarked, it would be interesting to evaluate these coefficients for QCD on the lattice.  This can be done because these coefficients
all have Kubo relations directly in terms of finite-temperature, Euclidean correlation functions of the sort which can be evaluated on the
lattice (without the need for analytic continuation).  This is possible precisely because these transport coefficients are thermodynamical
in nature.  Explicit expressions for the Kubo relations are found in the next section, see \Eq{kappa} to \Eq{lambda4}.  In Section
\ref{sec:latt}, we present a brief discussion of how these Kubo relations might be applied on the lattice.  In particular, we discuss
operator normalization and contact terms. Based on this discussion, we think the evaluation of $\kappa$ should be feasible with existing
techniques, at least for pure-glue QCD \cite{Schaefer}.  The evaluation of $\lambda_3$ and $\lambda_4$ may be prohibitively difficult since
computation of three-point functions are always much harder than two-point functions on lattice.

\section{Hydrodynamics}
\label{sec:hydro}

Hydrodynamics is a general theoretical framework for describing the
behavior of fluids locally near equilibrium.  It is organized as an
expansion in gradients of the fluid properties.  (For a recent review of
relativistic hydrodynamics see \cite{Romatschke,Teaney}).  At lowest (zero) order
in gradients, hydrodynamics is determined by equilibrium
thermodynamics.  The state of the fluid at each point is determined by
the values of all conserved charge densities.  For QCD, these are the
momentum density ${\cal P}^\mu$ associated with the stress tensor
$T^{\mu\nu}$ and the charge densities $Q_a$, $a=u,d,s,\ldots$ associated
with the conserved 4-currents $J^\mu_a$.  Assuming local equilibrium and
an equation of state for the pressure $P$ in terms of the energy density
and charge densities, $P=P(\epsilon,n_a)$, these currents can be
determined in terms of the conserved charge densities.  In practice one
uses a slightly more convenient set of variables;%
\footnote{Note that we are using the Landau-Lifshitz frame.}
the energy density and
flow 4-velocity, ${\cal P}^{\mu} = \epsilon u^\mu$ with
$g_{\mu\nu} u^\mu u^\nu=-1$ (where
$g_{\mu\nu}$ is the metric tensor, in flat space
$g_{\mu\nu} = \eta_{\mu\nu} = {\rm Diag}\:[-1,1,1,1]$)
and the number densities $n_a\equiv g_{\mu\nu} u^\mu J^\nu_a$.
In terms of these the stress tensor and current at lowest order are
\begin{eqnarray}
\label{T_zero}
T^{\mu\nu}(\epsilon,u,n) & = &
(\epsilon{+}P) u^\mu u^\nu + P g^{\mu\nu}
= \epsilon u^\mu u^\nu + P \Delta^{\mu\nu}
 \,, \\
J^\mu_a(\epsilon,u,n) & = & u^\mu n_a \,.
\label{J_zero}
\end{eqnarray}
Here $\Delta^{\mu\nu} \equiv g^{\mu\nu} + u^\mu u^\nu$ is a projection
operator onto the local spatial directions.  When $T^{\mu\nu}$ and
$J^\mu$ satisfy these expressions, then
stress conservation, $\nabla_\mu T^{\mu\nu}$ and current conservation,
$\nabla_\mu J^\mu_a$, close and completely determine the fluid
dynamics.  The relevant quantities at this order -- the pressure $P$ and
its various derivatives which give the entropy density $s$, quark number
susceptibilities $\chi_{ab}$, the speed of sound $c_s$, and so forth --
have been extensively studied on the lattice \cite{Lattice, Fodor}.

At first order in gradients
there are two independent terms one can add to the righthand side of
\Eq{T_zero}:
\bea
\label{T_one}
T^{\mu\nu} & = & \mbox{RHS of \Eq{T_zero} } -
\eta \sigma^{\mu\nu}
-\zeta \Delta^{\mu\nu} \nabla_\alpha u^\alpha \,,
\\
\sigma^{\mu\nu} & \equiv & \Delta^{\mu\alpha} \Delta^{\nu\beta}
 \left( \nabla_\alpha u_\beta + \nabla_\beta u_\alpha
   -\frac{2}{3} \Delta_{\alpha\beta} \nabla_\gamma u^\gamma \right) \,.
\eea
Here $\eta$, $\zeta$ are the shear and bulk viscosities
respectively. These new coefficients $\eta$, $\zeta$, and the
diffusion coefficient which can be added to \Eq{J_zero}, are all
dynamical quantities.  If the coefficients and the terms they
multiply are nonzero then entropy increases.  They can be
determined, via Kubo relations, from equilibrium correlation
functions of stress tensor or current operators, but the relations
involve evaluating these correlation functions at nonzero frequency,
which makes a {\sl direct} evaluation on the lattice impossible and
an {\sl indirect} evaluation at best very challenging
\cite{Aarts,Meyer2,Mikko2}.  However there is significant progress
in determining them from experiment \cite{Expr}.

At second order there are a host of terms which can be added.  The
situation improves somewhat if we assume that charge densities are
small, so $J^\mu_a$ terms can be neglected.  If in addition we assume
that the theory under consideration is conformally invariant, then there
are 5 additional terms which must be included \cite{BRSSS}.  However,
since we are interested in QCD at finite coupling and potentially in
making contact with the lattice, we cannot assume conformal invariance.
In this case, after reducing the number of terms by applying equations
of motion and other inter-relations, there are 15 independent terms which
appear at second order, which have been enumerated by Romatschke
\cite{Rom_entropy}.
To write these terms explicitly it is convenient to introduce
the vorticity tensor $\Omega^{\mu\nu}$,
\be
2 \Omega^{\mu\nu} \equiv \Delta^{\mu\alpha} \Delta^{\nu\beta}
  \left( \nabla_\alpha u_\beta - \nabla_\beta u_\alpha \right) \,,
\ee
as well as the curvature tensor $R^{\mu\nu\alpha\beta}$ and
Ricci tensor $R^{\mu\nu} = {{R^{\mu}}_\alpha}^{\nu\alpha}$ and
scalar $R = {R^\mu}_\mu$.  And we will write
\be
R^{\mu\langle \nu
\alpha \rangle \beta} \equiv \half R^{\mu \kappa \sigma \beta}
\left(
 \Delta^{\nu}_\kappa \Delta^{\alpha}_{\sigma}
 +\Delta^{\nu}_\sigma \Delta^{\alpha}_{\kappa}
-\frac{2}{3} \Delta^{\nu\alpha} \Delta_{\kappa\sigma}
 \right)
\ee
and similarly for $R^{\langle \mu\nu \rangle}$.  That is, the indices enclosed
in angle brackets are space-projected, symmetrized, and trace-subtracted.
Using all of this notation, the possible second-order terms, according to
Romatschke \cite{Rom_entropy}, are
\bea
\label{T_order2}
T^{\mu\nu} & = & \mbox{\Eq{T_one}}
  + \eta \tau_\pi \left( u\cdot \nabla \sigma^{\mu\nu}
                         + \frac{\nabla \cdot u}{3} \sigma^{\mu\nu} \right)
\nonumber \\ &&
  + \kappa \left( R^{\langle \mu\nu \rangle}
                 - 2 u_\alpha u_\beta R^{\alpha\langle \mu\nu \rangle \beta}
                \right)
\nonumber \\ &&
  + \lambda_1 {\sigma_\lambda}^{\langle \mu} \sigma^{\nu\rangle \lambda}
  + \lambda_2 {\sigma_\lambda}^{\langle \mu} \Omega^{\nu\rangle \lambda}
  - \lambda_3 {\Omega_\lambda}^{\langle \mu} \Omega^{\nu\rangle \lambda}
\nonumber \\ &&
  + \eta \tau^*_\pi \frac{\nabla \cdot u}{3} \sigma^{\mu\nu}
  + \lambda_4 \nabla^{\langle \mu} \ln s \nabla^{\nu\rangle} \ln s
  + 2 \kappa^* u_\alpha u_\beta R^{\alpha \langle \mu\nu \rangle \beta}
\nonumber \\ &&
 + \Delta^{\mu\nu} \left( - \zeta \tau_\Pi u\cdot \nabla \nabla \cdot u
   +\xi_1 \sigma^{\alpha\beta} \sigma_{\alpha\beta}
   +\xi_2 (\nabla \cdot u)^2
\right. \nonumber \\ && \left. \hspace{3.5em}
   +\xi_4 \nabla_{\alpha\perp} \ln s \nabla^{\alpha}_{\perp} \ln s
   +\xi_3 \Omega^{\alpha\beta} \Omega_{\alpha\beta}
   +\xi_5 R
   +\xi_6 u^\alpha u^\beta R_{\alpha\beta}
  \right) \,.
\eea

There are several ways to categorize these terms.  Some are only
relevant in curved space; $\kappa$, $\kappa^{*}$, $\xi_5$ and
$\xi_6$. The others are relevant in flat or curved space.  (Even
though $\kappa$ {\it etc}.\ only play a role in curved space, they
mix with the other terms when we find Kubo relations in
Eqs.~(\ref{kappa}-\ref{lambda4}), so they should generally be
considered anyway \cite{BRSSS}.) We can also divide the terms into
linear and nonlinear terms.  Linear terms affect small fluctuations
and can, for instance, influence their dispersion; nonlinear terms
are only relevant at second order in small fluctuations about
equilibrium and flat space.  The linear terms are $\tau_\pi$,
$\kappa$, $\kappa^*$, $\tau_\Pi$, $\xi_5$, and $\xi_6$. The other
terms, $\lambda_{1\ldots 4}$, $\tau_\pi^*$, $\xi_{1\ldots 4}$ are
nonlinear.

We can also group these terms into those which are thermodynamical in
nature, and those which are dynamical.  We call a term thermodynamical
if it can give a nonzero contribution to $T^{\mu\nu}$ when the geometry
and density matrix are fully time-independent and the system is therefore in
equilibrium.  No term involving the shear tensor $\sigma_{\mu\nu}$
is thermodynamical because a system under shear flow is changing with
time and is producing entropy.  In nonconformal theories, the same is
true of bulk flow
$\nabla \cdot u$.  However, it is completely consistent to have
a system which is in equilibrium in a curved (but time-independent)
geometry.  Similarly, a time-independent but space-varying $g_{00}$
(gravitational potential) makes $\nabla_{\mu\perp} s$ nonzero without any
departure from equilibrium.
Similarly, it is possible (in a curved geometry) to establish
persistent vorticity which is sustained forever.%
\footnote{%
    For instance, consider a spacetime which is
    ${\cal S}^2 \times {\cal R}^2$, with time in one of the flat
    directions.  The fluid can spin about the equator of the
    ${\cal S}^2$ and this flow will persist forever, and will therefore
    reach equilibrium.}
The system will be fully in equilibrium in the presence of this
vorticity.  Hence, the coefficients $\kappa$, $\kappa^*$,
$\lambda_3$, $\lambda_4$, $\xi_3$, $\xi_4$, $\xi_5$, $\xi_6$ represent
thermodynamical quantities.

In Ref.~\cite{MooreSohrabi} we showed how to derive Kubo relations
for second-order hydro coefficients.  There we did so only for
conformal theories, but it is straightforward to do so for nonconformal
theories as well.  Doing so, we find that the Kubo relations for
the thermodynamical coefficients can all be expressed in terms of
retarded correlation functions evaluated directly at zero frequency.  Up
to powers of $i$, zero-frequency retarded correlators equal
zero-frequency Euclidean correlators.  In fact, we can derive (Kubo)
relations between the thermodynamic coefficients and Euclidean
correlators by working directly in Euclidean space.
In particular, defining the Euclidean $n$-point function as
\begin{eqnarray} \label{action_E} G_{\rm E}^{\mu_1\nu_1
   \ldots \mu_n \nu_n}(p_{1},.\,.\,,p_{n-1},-p_1{-}.\,.\,{-}p_{n-1})
 &\equiv& \left. \int d^4 x_1 \ldots d^4 x_{n-1}
    e^{-i(p_1\cdot x_1+\ldots +p_{n-1}\cdot x_{n-1})}\right.{}
\nonumber\\
  & & {}\left. \times
\frac{2^n \; \partial^n \ln {\rm Z}}
  {\partial g_{\mu_1\nu_1}(x_1) \ldots \partial g_{\mu_n \nu_n}(0)}
\right|_{g_{\mu\nu} = \delta_{\mu\nu}}{}
\end{eqnarray}
 with
\begin{eqnarray}
Z\left[g_{\mu\nu}\right]=\int\mathcal{D}\phi  \; \exp
\left\{-S_{\rm E}\left[\phi,g_{\mu\nu}\right]\right\} \,,
\end{eqnarray}
we find
\bea
\label{kappa}
\kappa & \!=\! & \lim_{k_{z} \rightarrow
0}
    \frac{\partial^2}{\partial k^2_{z}}G^{xy,xy}_{E}(k)|_{k_{0}=0}\,,
\\ \label{lambda3}
\lambda_3 &\!=\!& 2\kappa^{*} - 4 \lim_{p^z,q^z \rightarrow 0}
             \frac{\partial^2}{\partial p_{z}\partial q_{z}}
             G^{xt,yt,xy}_{E}(p,q)|_{p_{0}, q_{0}=0} \,,
\\ \label{lambda4}
\lambda_{4}&\!=\!& -2\kappa^{*}+\kappa
-\frac{c^{4}_{s}}{2}\lim_{p^{x}, q^{y} \rightarrow 0} \frac{\partial^2}{\partial p_{x}\partial
q_{y}}G_{E}^{tt,tt,xy}(p,q)|_{p_{0}, q_{0}=0} \,.
\eea
The remaining transport
 coefficients, including $\kappa^*$ which appears above, are
not independent but are determined in terms of these three via five
 independent conditions.  Two conditions were found by Romatschke
\cite{Romatschke}, by demanding that the entropy current have non-negative
 divergence.  His calculation was limited to second order in
gradients; but a treatment to third order in gradients by Bhattacharyya
 \cite{Bhattacharyya} and Jensen {\it et al} \cite{Jensen} found three more constraints on second order
transport coefficients. For an interesting physical interpretation of these constraints, see \cite{Minwalla2}.

The five constraints found by Bhattacharyya, in our notation%
\footnote{Labeling the coefficients of \cite{Minwalla2} with a prime,
  the relations between their coefficients $\kappa_1'$,
$\kappa_2'$, $\lambda_3'$, $\lambda_4'$, $\zeta_2'$, $\zeta_3'$,
$\xi_3'$, and $\xi_4'$ and our coefficients are:
$T\kappa_{1}'\!=\!\kappa$, $T\kappa_{2}'\!=\!2\kappa-2\kappa^{*}$,
$-T\lambda^{'}_{3}\!=\!\lambda_{3}$,
$c^4_{s}T\lambda^{'}_{4}\!=\!\lambda_{4}$, $T\zeta_{2}'\!=\!\xi_{5}$,
$T\zeta_{3}'\!=\!\xi_{6}$, $-T\xi^{'}_{3}\!=\!\xi_{3}$ and
$c^{4}_{s}T\xi^{'}_{4}\!=\!\xi_{4}$.
Moreover, unlike \cite{Minwalla2} our convention for
$R^{\rho\sigma\mu\nu}$ is $R^{\rho}_{\
\sigma\mu\nu}=\partial_{\mu}\Gamma^{\rho}_{\nu\sigma}-\partial_{\nu}\Gamma^{\rho}_{\mu\sigma}+
\Gamma^{\rho}_{\mu\lambda}\Gamma^{\lambda}_{\nu\sigma}
-\Gamma^{\rho}_{\nu\lambda}\Gamma^{\lambda}_{\mu\sigma}$. },
are
\bea
\label{dummykappa}\kappa^{*}&=&\kappa-\frac{T}{2}\frac{d\kappa}{dT}\,,
\\
\label{dummxi_{5}}\xi_{5}&=&
\frac{1}{2}\left(c^{2}_{s}T\frac{d\kappa}{dT}-c^2_{s}\kappa-\frac{\kappa}{3}\right)\,,
\\
\label{dummyxi_{6}}\xi_{6}&=&c^2_{s}\left(3T\frac{d\kappa}{dT}
-2T\frac{d\kappa^{*}}{dT}+2\kappa^{*}-3\kappa\right)-\kappa+\frac{4\kappa^{*}}{3}
+\frac{\lambda_{4}}{c^2_{s}}\,,
\\\nonumber
\label{dummyxi_{3}} \xi_{3}&=&\frac{3c^2_{s}T}{2}\left(\frac{d\kappa^{*}}{dT}-
 \frac{d\kappa}{d T}\right)+\frac{3\left(c^2_{s}-1\right)}{2}(\kappa^{*}-\kappa)
 -\frac{\lambda_{4}}{c^2_{s}}
 +\frac{1}{4}\left(c^2_{s}T\frac{d\lambda_{3}}{dT}-3c^2_{s}\lambda_{3}
 +\frac{\lambda_{3}}{3}\right),
\nonumber\\
\\
\label{dummyxi_{4}} \xi_{4}&=&-\frac{\lambda_{4}}{6} -
 \frac{c^2_{s}}{2}\left(\lambda_{4}
 +T\frac{d\lambda_{4}}{d
 T}\right)+c^4_{s}\left(1-3c^2_{s}\right)\left(T\frac{d\kappa}{dT}
 -T\frac{d\kappa^{*}}{dT}+\kappa^{*}-\kappa\right)
 \nonumber\\&&-c^{6}_{s}T^{3}\frac{d^2}{dT^2}\left(\frac{\kappa-\kappa^{*}}{T}\right)\,.
\eea
We take these constraints to determine all other coefficients in
 terms of $\kappa$, $\lambda_3$, and $\lambda_4$. In Appendix
\ref{appendixA} we give a detailed derivation of Eqs.(\ref{kappa}--\ref{lambda4}),
and we find Kubo relations for the dependent transport
coefficients mentioned in Eqs.~(\ref{dummykappa}--\ref{dummyxi_{4}}) for completeness.

Euclidean correlation functions have well behaved perturbative
expansions at finite temperature (at least at low order), therefore
it should be possible to evaluate these correlators perturbatively
in a weakly coupled theory. We present the derivation at lowest
order in a general massless theory for particles of spin zero, half
and one in the next section.


\section{Evaluation at Weak Coupling}
\label{sec:weakcoupling}

To carry out the calculation of these transport coefficients, first
we have to clarify the nature of the correlation functions that are
derived by differentiating the curved space partition function. The
definition for the $n$-point Green function established in
\Eq{action_E} involves multiple derivatives acting on the energy
functional.  Each derivative can pull down a factor of $-2\partial
{\cal L_{\rm E}}/\partial g_{\mu\nu}= T^{\mu\nu}$, giving a
conventional $n$-point stress-tensor correlator; but the
$g_{\mu\nu}$ derivatives can also act on $T^{\alpha\beta}$ factors
pulled down by previous $g_{\alpha\beta}$ derivatives, leading to
contact terms.  (In intermediate steps our $T^{\mu\nu}$ is really the
stress tensor density $\sqrt{g}\,T^{\mu\nu}$; the distinction is
irrelevant in final expressions since in the end we evaluate correlators
in flat space.)  So in terms of the usual $n$-point stress tensor
correlators, $G_{E}^{\mu\nu,\ldots,\alpha\beta}$ defined in
\Eq{action_E} is
\bea
\label{two-point}
G^{\mu\nu,\alpha\beta}_{E}(0,x)&\!=\!&\left. \left\langle
    T^{\mu\nu}(0)T^{\alpha\beta}(x)\right\rangle
        \right|_{g_{\mu\nu}=\eta_{\mu\nu}}
+2\left. \left\langle \frac{\partial T^{\mu\nu}(0)}
      {\partial g_{\alpha\beta}(x)}\right\rangle
          \right|_{g_{\mu\nu}=\eta_{\mu\nu}}\,,\\
\label{three-point}
G^{\mu\nu,\alpha\beta,\gamma\rho}_{E}(0,x,y)&\!=\!&
  \left. \left\langle T^{\mu\nu}(0)T^{\alpha\beta}(x)
                 T^{\gamma\rho}(y)\right\rangle\right|_{g_{\mu\nu}
  =\eta_{\mu\nu}} +2\left. \left\langle
  \frac{\partial T^{\mu\nu}(0)}
  {\partial g_{\alpha\beta}(x)}T^{\gamma\rho}(y)\right\rangle
        \right|_{g_{\mu\nu} =\eta_{\mu\nu}}
\\\nonumber&&
+2\left. \left\langle \frac{\partial T^{\mu\nu}(0)}{\partial
g_{\gamma\rho}(y)}T^{\alpha\beta}(x)
         \right\rangle\right|_{g_{\mu\nu} = \eta_{\mu\nu}}
  +2\left. \left\langle T^{\mu\nu}(0)
       \frac{\partial T^{\alpha\beta}(x)}{\partial g_{\gamma\rho}(y)}
  \right\rangle\right|_{g_{\mu\nu} = \eta_{\mu\nu}}
 \\\nonumber&&
 +4\left. \left\langle
\frac{\partial^2 T^{\mu\nu}(0)}{\partial g_{\alpha\beta}(x)
   \partial g_{\gamma\rho}(y)}\right\rangle
   \right|_{g_{\mu\nu}= \eta_{\mu\nu}}\,.
\eea
The terms involving derivatives of the stress tensor are called contact terms,
and are discussed in some detail in Ref.\ \cite{RomatschkeSon}.  Since
$\partial T^{\mu\nu}(x)/\partial g_{\alpha\beta}(y)
  \propto \delta^4(x-y)$
they have very simple momentum dependence.  In particular, the contact
term in $G^{\mu\nu,\alpha\beta}_{E}(x)$ is $\propto \delta^4(x)$; so its
contribution to $G^{\mu\nu,\alpha\beta}_{E}(k)$ is $k$-independent.
Therefore it does not contribute to \Eq{kappa}.

Now consider the four contact terms in \Eq{three-point} and their
contribution to \Eq{lambda3}.  Defining
\be
2\frac{\partial T^{\mu\nu}(x)}{\partial g_{\alpha\beta}(y)}
\equiv X^{\mu\nu\alpha\beta} \delta^4(x-y) \,,
\label{Contactterm}
\ee
we find three $X$-type contact terms,
involving $\delta^4(x)$, $\delta^4(y)$, and $\delta^4(x-y)$
respectively.  The first gives a contribution which is independent of
$p$ and so does not contribute to \Eq{lambda3}; similarly the second is
independent of $q$ and also does not contribute.  But the third term
does contribute;
\begin{eqnarray}
\label{Lambda3} \lambda_{3}& = & - 4\lim_{p_{z},q_{z} \rightarrow
0}\partial_{p_z}
\partial_{q_z}
   \left\langle T^{xt}(p) T^{yt}(q) T^{xy}(-p-q) \right\rangle
\nonumber\\ && - 4\lim_{p_{z},q_{z} \rightarrow 0}\partial_{p_z}
\partial_{q_z}
   \left\langle X^{xtyt}(p+q) T^{xy}(-p-q) \right\rangle\, .
\end{eqnarray}

In order to calculate these transport coefficients for a generic
field theory, we need to find the explicit form of both the stress
tensor $T^{\mu\nu}$ and of the contact term $X^{\mu\nu\alpha\beta}$ by
differentiating the action
\begin{eqnarray}\label{action-Eucl}
S=\int d^4x
\left(\mathcal{L}_{\rm scalar}+\mathcal{L}_{\rm spinor}
 +\mathcal{L}_{\rm vector}\right)
\end{eqnarray}
with respect to the metric.  Since we only attempt a leading-order
calculation here, it is sufficient to consider the free-theory action in
curved space,
\bea
\label{Lscalar}
\mathcal{L}_{\rm scalar}
&=&\frac{\sqrt{g}}{2}g^{\mu\nu}\partial_{\nu}\phi\partial_{\mu}\phi\,,\\
\label{Lspinor}
\mathcal{L}_{\rm spinor}&=& |e| \: \bar{\psi}\gamma^{c}e^{\lambda}_{c}
  \left(\partial_{\lambda}
  +\frac{1}{2}G^{ab}\omega^{ab}_{\lambda}\right)\psi\,,\\
\label{Lvector}
\mathcal{L}_{\rm vector}&=&\frac{\sqrt{g}}{4}
   F_{\mu\nu}F_{\rho\tau}g^{\mu\rho}g^{\nu\tau}\,.
\eea
Here $g^{\mu\nu}$ is the inverse of $g_{\mu\nu}$,
$F_{\mu\nu} = \partial_\mu A_\nu - \partial_\nu A_\mu$ is the field
strength tensor, $e_\mu^a$ is the vierbein related to $g_{\mu\nu}$ by
$\eta_{ab} e^{a}_{\mu}e^{b}_{\nu}=g_{\mu\nu}$ and
$|e|=\det\left(e^{a}_{\mu}\right)$ is its determinant.  Finally,
$\omega^{ab}_{\lambda}$ is the
spin connection and
$G^{ab}=\frac{1}{4}\left[\gamma^{a},\gamma^{b}\right]$.

Actually, more generally the scalar Lagrangian density should read
\be
\label{Lscalar2}
\mathcal{L}_{\rm scalar}=\frac{\sqrt{g}}{2}
   \left(g^{\mu\nu}\partial_{\nu}\phi\partial_{\mu}\phi
   -\xi R \phi^2\right) \,,
\ee
where $\xi$ is a dimensionless constant and $R$ is the Ricci scalar
introduced earlier.  The action is conformal for the choice
$\xi=\frac{1}{6}$ and is called {\sl minimally coupled} if
$\xi=0$ \cite{Parker}.  We will consider  general $\xi$, but in the end
our results for $\kappa,\lambda_3$ are $\xi$ independent.

Through the reminder of this section we will compute $\kappa$ and $\lambda_{3}$ using \Eq{kappa} and \Eq{Lambda3}, applying the action in
\Eq{action-Eucl}. We also note that at the leading order calculation, $\lambda_{4}=0$ due to conformal symmetry. Since the effect of more
degrees of freedom $N_{0}$, $N_{1/2}$ and $N_{1}$ at this level is multiplicative, they can be counted in the final result respectively.


\subsection{Scalars}

\label{sec:scalars}

In carrying out the variation of \Eq{Lscalar2} with respect to
$g_{\mu\nu}$, we must consider the explicit dependence and the
implicit dependence via the Ricci scalar $R$.  The resulting stress
tensor is
\be
\label{Tscalar}
T^{\mu\nu} = \left( (1-2\xi)
g^{\mu\alpha}g^{\nu\beta}
        +\frac{4\xi - 1}{2} g^{\mu\nu}g^{\alpha\beta}\right)
     \partial_{\alpha}\phi\partial_{\beta}\phi
 +2\xi \left(g^{\mu\nu}g^{\alpha\beta}{-} g^{\mu\alpha}g^{\nu\beta}\right)
     \phi\partial_{\alpha}\partial_{\beta}\phi\,,
\ee
plus terms which vanish in flat space.

\begin{figure}
\centerline{
 \begin{picture}(120,80)
   \thicklines
   \put(30,40){\insertion}
   \put(90,40){\insertion}
   \put(60,44){\oval(60,40)[t]}
   \put(60,36){\oval(60,40)[b]}
   \thinlines
   \put(8,40){\vector(1,0){10}}
   \put(102,40){\vector(1,0){10}}
   \put(55,70){\vector(1,0){10}}
   \put(65,10){\vector(-1,0){10}}
   \put(0,37){$k$}
   \put(115,37){$k$}
   \put(51,75){$p{+}k$}
   \put(56,0){$p$}
 \end{picture}}
\caption{\small  Leading order scalar diagram contributing to
 $\langle T^{xy}(-k) T^{xy}(k)\rangle$, necessary for evaluation of
 $\kappa$.  The crosses are $T$ insertions, the solid lines are scalar
 propagators, and the arrows indicate the momenta flowing on lines and
 entering or leaving $T$ insertions.  \label{scalarkappa}}
\end{figure}
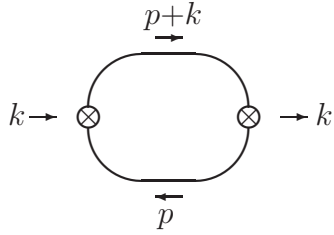

With $T^{\mu\nu}$ in hand, we can compute the scalar contribution to
$\kappa$. The lowest order diagram is shown in Figure
\ref{scalarkappa}.  The momentum $k$ enters at one $T^{xy}$ insertion
and exits at the other, so the scalar propagators carry different
momenta, $p$ and $p{+}k \equiv q$. $T^{xy}$ inserted between these lines
obeys the Feynman rule (note directions of momentum flow)
\be
\begin{picture}(60,10)
\thicklines
\put(0,-7){ \begin{picture}(60,10)
\put(0,10){\line(1,0){20}}
\put(24,10){\insertion}
\put(28,10){\line(1,0){20}}
\put(1,0){$p$}
\put(31,0){$q$}
\thinlines
\put(21,17){$T^{xy}$}
\put(8,2){\vector(1,0){10}}
\put(38,2){\vector(1,0){10}}
\end{picture}}\end{picture}
 = (1-2\xi)\Big(p^x q^y + q^x p^y\Big)
   + 2\xi \Big( p^x p^y + q^x q^y \Big) \,.
\ee
Since we are only differentiating with respect to $k_z$, we may set
$k_x=0=k_y$ from the outset, in which case $p^x=q^x$ and $p^y = q^y$.
Therefore the $\xi$ terms cancel%
\footnote%
    {The same will not happen if we compute non-conformal coefficients.}
and the diagram evaluates to
\begin{eqnarray}
\kappa = \partial^{2}_{k_z} \langle T^{xy}(-k) T^{xy}(k)\rangle
 & = & \left. \frac{1}{2}\partial^{2}_{k_{z}}\ \sumint_{p}
     \frac{(2 p^{x}p^{y})^2}{p^2(p{+}k)^2} \right|_{k=0}
\nonumber\\
 & = & -4T\sumint_{p} \left(\frac{1}{p^6}
                          -\frac{4p^{2}_{z}}{p^8}\right)
   p^{2}_{x}p^{2}_{y}
 = -\frac{T^{2}}{72} \,,
\end{eqnarray}
where $\half$ is the symmetry factor of the diagram,
and the integration-summation symbol is defined as
\be
\sumint_{p} = T \hspace{-0.8em} \sum_{p^0=2\pi nT}
              \int\frac{d^3\vec{p}}{(2\pi)^3}
\ee
and $n$ runs over the integers.  In evaluating this and related
sum-integrals we use the result
\be
\label{sumint}
\sumint_p \frac{(\vec{p}^{\,2})^n}{(p^{2})^{n+1}}
 = \frac{(2n+1)!}{2^{2n} (n!)^2} \: \frac{T^2}{12} \,.
\ee
Expressions with powers of $(p^0)^2$ in the numerator can be handled by
rewriting $(p^0)^2 = p^2 - \vec{p}^{\,2}$ and using this relation
repeatedly; for instance,
$\sumintsmall_p \frac{(p^0)^6}{(p^2)^4}
  = \frac{-1}{16}\: \frac{T^2}{12}$.  We handle
$p_x^2 p_y^2 p_z^2$ by angular averaging,
$\langle p_x^2 p_y^2 p_z^2 \rangle_{\rm angle} = \vec{p}^{\,6}/105$.

Our leading-order result for $\kappa$
agrees with the result in Ref.\ \cite{RomatschkeSon}. The above result
shows that the weak coupling expansion of $\kappa$ starts at
$\alpha^{0}$.

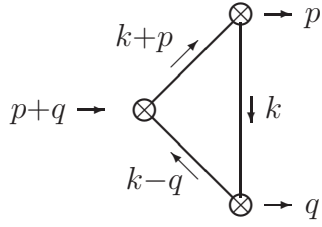
\begin{figure}
\centerline{
 \begin{picture}(125,100)
   \thinlines
   \put(5,47){$p{+}q$}
   \put(30,50){\vector(1,0){10}}
   \put(100,86){\vector(1,0){10}}
   \put(100,14){\vector(1,0){10}}
   \put(115,83){$p$}
   \put(115,11){$q$}
   \put(95,55){\vector(0,-1){10}}
   \put(65,66){\vector(1,1){10}}
   \put(75,24){\vector(-1,1){10}}
   \put(100,47){$k$}
   \put(48,20){$k{-}q$}
   \put(44,74){$k{+}p$}
   \thicklines
   \put(55,50){\insertion}
   \put(58,53){\line(1,1){30}}
   \put(91,86){\insertion}
   \put(91,83){\line(0,-1){66}}
   \put(91,14){\insertion}
   \put(58,47){\line(1,-1){30}}
   \thicklines
 \end{picture}}
\caption{Three-point correlation function
$\langle T^{xt}(p) T^{yt}(q) T^{xy}(-p-q) \rangle$
that contributes to the Kubo formula of
$\lambda_{3}$; the leftmost vertex is $T^{xy}$, the other vertices are
$T^{xt}$ and $T^{yt}$. \label{scalarlam3diagram1}}
\end{figure}

Now we turn to the computation of $\lambda_{3}$. The first term
appearing in \Eq{Lambda3} is represented by the diagram shown in Figure
\ref{scalarlam3diagram1}.  Once again, $p,q$ only need have
nonvanishing $z$-components, but the $T$ operators only return $x,y,0$
components, which simplifies the evaluation of the diagram and ensures
that the result is $\xi$ independent.  The diagram evaluates to
\begin{eqnarray}
\label{TriangleScalar}
& & -4\partial_{p_z} \partial_{q_z}
   \left\langle T^{xt}(p) T^{yt}(q) T^{xy}(-p-q) \right\rangle
\nonumber\\ & = & - \left.
 4\partial_{p_{z}}\partial_{q_{z}}\sumint_{k}
   \frac{(2k^x k^y)
         (2k^t k^x)
         (2k^t k^y)}
       {(k{+}p)^2 (k{-}q)^2 k^2} \right|_{p,q=0}
\nonumber\\ & = &
 128\sumint_{k}\frac{k^2_{t}\
k^2_{x}\ k^2_{y}\ k^2_{z}}{k^{10}}=-\frac{T^2}{36} \,.
\end{eqnarray}
For the contact term in \Eq{Lambda3}, we need to calculate $X^{ytxt}$ as
defined in \Eq{Contactterm}. Variation of \Eq{Tscalar} with
respect to $g_{xt}$ gives gives
\begin{eqnarray}
X^{ytxt} =  - \partial^{y}\phi\partial^{x}\phi \,,
\end{eqnarray}
in flat space and for $\xi=0$. In addition there are terms proportional to
$\xi$, but they again always involve the combination
$\partial_\alpha \partial_\beta \phi^2$.  These terms do not contribute to
the correlation function we need for the same reason the
$\xi$-proportional terms above did not contribute; the incoming and outgoing
 momenta equal for the components which make up the indices of
$X^{txty}$.  Therefore the result is again $\xi$ independent. The contribution
 from the $\langle TX\rangle$ correlator to $\lambda_3$ is
\begin{eqnarray}
\label{ContactScalar}
-4\partial_{p_z}\partial_{q_z}
      \left\langle X^{xtyt}(p+q) T^{xy}(-p-q) \right\rangle
 & = & 2\partial_{p_{z}}\partial_{q_{z}}\sumint_{k}
 \frac{(2k_x k_y)^2}{(k{-}p)^2 (k{+}q)^2}
\nonumber \\ & = &
-32 \sumint_k \frac{k_x^2 k_y^2 k_z^2}{(k^2)^4}
=-\frac{T^2}{18} \,.
\end{eqnarray}
This diagram is actually the same as the diagram which determines
$\kappa$; shifting the integration variable used above by $p$, the
integral becomes the same one needed in evaluating $\kappa$ except for
the overall factor of $4$.
Summing up \Eq{TriangleScalar} and \Eq{ContactScalar}, we get
\begin{eqnarray}
\lambda_{3}=-\frac{T^2}{12}\text{\quad 1 real scalar field} \,.
\end{eqnarray}
We follow a similar calculation for gauge and fermion
fields in the next sections.
\subsection{Gauge fields}

The gauge field stress tensor derived by $g_{\mu\nu}$ variation using
the gauge field action, \Eq{Lvector}, is
\begin{eqnarray}
\label{gaugetensor} T^{\mu\nu}=F^{\mu\alpha}F^{\nu}_{\ \alpha}
          -\frac{1}{4}g^{\mu\nu}F^{\alpha\beta}F_{\alpha\beta}\,,
\end{eqnarray}
and from the above relation, we derive the Feynman rule for the
vertex,
\bea
\label{GaugeTFeynman}
\begin{picture}(90,10)
  \thicklines
  \put(-10,0){$\mu,a$}
  \multiput(14,2)(8,0){3}{\oval(4,4)[t]}
  \multiput(18,2)(8,0){2}{\oval(4,4)[b]}
  \put(36,2){\insertion}
  \put(32,9.5){$T^{\alpha\beta}$}
  \multiput(42,2)(8,0){3}{\oval(4,4)[t]}
  \multiput(46,2)(8,0){2}{\oval(4,4)[b]}
  \put(62,0){$\nu,b$}
  \put(22,-8){\vector(1,0){10}}
  \put(12,-10){$p$}
  \put(52,-8){\vector(1,0){10}}
  \put(43,-10){$k$}
\end{picture}
&&
\delta_{ab} \Big( (p_\alpha g_{\mu\gamma} - p_\gamma g_{\mu \alpha})
                   (k_\beta  g_\nu^\gamma -  k^\gamma g_{\nu \beta} )
             +(\mu\leftrightarrow \nu)
 \nonumber \\ && \;\quad {}
             -g_{\alpha\beta}
                 (p\cdot k \: g_{\mu\nu} - k_\mu p_\nu) \Big)\,.
\eea
The expression for $X^{\mu\nu\alpha\beta}$ is rather long,
but for the case of interest, $X^{txty}$, it is quite simple:
\be
X^{txty} = -F^{xz} F^{y}{}_z
\ee
leading to the Feynman rule
\be
\begin{picture}(90,10)
  \thicklines
  \put(-10,0){$\mu,a$}
  \multiput(14,2)(8,0){3}{\oval(4,4)[t]}
  \multiput(18,2)(8,0){2}{\oval(4,4)[b]}
  \put(36,2){\insertion}
  \put(29,9){$X^{0x0y}$}
  \multiput(42,2)(8,0){3}{\oval(4,4)[t]}
  \multiput(46,2)(8,0){2}{\oval(4,4)[b]}
  \put(62,0){$\nu,b$}
  \put(22,-8){\vector(1,0){10}}
  \put(12,-10){$p$}
  \put(52,-8){\vector(1,0){10}}
  \put(43,-10){$k$}
\end{picture} \;\; {}-
\delta_{ab} \Big( (p_x g_{\mu z} - p_z g_{\mu x})
                  (k_y g_{\nu z} - k_z g_{\nu y})
                  + (x\leftrightarrow y) \Big) \,.
\ee

The calculation of $\kappa$ and $\lambda_3$ then proceeds via the same
diagrams as in the scalar case, but with these somewhat more complicated
Feynman rules for the vertices, and with gauge propagators.  Note that,
because $T^{\mu\nu}$ is built from field strengths, it applies a
transverse projector onto the incoming gauge field index; contracting
\Eq{GaugeTFeynman} with $p_\mu$ or $k_\nu$ gives zero.  Therefore the
result is gauge parameter independent within covariant gauges (and
all linear gauges).  After significant algebra we find
that
\begin{eqnarray}
\kappa=\frac{T^2}{18} \text{\  for a single color}\,,
\end{eqnarray}
while the two diagrams contributing to $\lambda_3$ give
\be
\label{Trianglegauge}
-4\partial_{p_z} \partial_{q_z}
   \left\langle T^{xt}(p) T^{yt}(q) T^{xy}(-p-q) \right\rangle
 = \frac{2T^2}{9}
\ee
and
\be
\label{Contactgauge}
-4\partial_{p_z}\partial_{q_z}
  \left\langle X^{xtyt}(p+q) T^{xy}(-p-q) \right\rangle
 = \frac{T^2}{9} \,.
\ee
Therefore, the gauge field contribution to $\lambda_3$ is
\begin{eqnarray}
\lambda_{3}= \frac{T^2}{3}\text{\quad for a single color} \,.
\end{eqnarray}
Since each color possesses two spin states, we need to divide these
results by 2 to get expressions per degree of freedom.


\subsection{Fermions}

The treatment of fermions in curved space requires the introduction
of the vierbein (also called the frame vector or tetrad) $e^a_\mu$
(for a review and a treatment of their application to the stress
tensor see Ref.\ \cite{DeWitt}).  The vierbein relates a local
orthonormal coordinate system on the tangent space, with indices $a$
and metric $\eta_{ab}$ (which is $\delta_{ab}$ in Euclidean space)
to the metric, via
\be
\label{e_and_g}
\eta_{ab} e^a_\mu e^b_{\nu} =
g_{\mu\nu} \,;
\ee
in a sense it is the square root of the metric.
The Dirac action is $e_a^\mu \bar\psi \gamma^a \nabla_\mu \psi$,
where the action of $\nabla_\mu$ on a spinorial object is determined
by the spin connection $\omega^{ab}_\mu$:
\be
\nabla_{\mu}\psi=\partial_{\mu}\psi+\frac{1}{2}G_{[ab]}
\omega^{ab}_{\mu} \psi
\ee
where $G_{[ab]}=\frac{1}{4} \left[\gamma_{a},\gamma_{b}\right]$, and the
spin connection is related to the vierbein via
\be
\label{Identity}
\omega^{ab}_{\mu}=\frac{1}{2} e^{a\nu}(\partial_{\mu}e^{b}_{\nu}
-\partial_{\nu}e^{b}_{\mu})
-\frac{1}{2}e^{b\nu}(\partial_{\mu}e^{a}_{\nu}
-\partial_{\nu}e^{a}_{\mu})
+\frac{1}{2}e^{a\nu}e^{b\sigma}(\partial_{\sigma}e^{c}_{\nu}
-\partial_{\nu}e^{c}_{\sigma})e_{c\mu}\,.
\ee
Because the action
depends on the local frame components, the stress-tensor for
fermions cannot be obtained by functional differentiation with
respect to the metric tensor; instead one must use the more general
expression
$T^{\mu\nu}(x) = e_{a}^{\nu} \frac{\partial Z}{\partial
e_{a\mu}(x)}$,
which reduces to
$T^{\mu\nu}(x) = 2 \frac{\partial Z}
 {\partial g_{\mu\nu}(x)}$
for any terms which depend only on $g_{\mu\nu}$ because of
\Eq{e_and_g}.  Applying this relation to the fermionic action, noting
that
$\frac{\delta e^{a\mu}}{\partial e_{b\nu}} = -\eta^{ab} g^{\mu\nu}$
(since $g^{\mu\nu}$ is the inverse of $g_{\mu\nu}$; alternatively,
because the variation of $g^{\mu\nu} g_{\mu\nu}$ should vanish), and
specializing to the non-diagonal entries in $T^{\mu\nu}$, after
some work one obtains \cite{DeWitt}
\be
\label{StressTensorFermions}
T^{\mu\nu}=\frac{1}{4}\left(
    \bar{\psi}\gamma^{\mu}\: \nabla^{\nu}\psi
   -\nabla^{\mu}\bar{\psi}\: \gamma^{\nu}\psi
   +\bar{\psi}\gamma^{\nu}\: \nabla^{\mu}\psi
   -\nabla^{\nu}\bar{\psi}\: \gamma^{\mu}\psi\right) \,.
\ee
The relevant Feynman rule is
\be
\begin{picture}(60,10)
\thicklines
\put(0,-7){ \begin{picture}(60,10)
\put(0,10){\line(1,0){20}}
\put(7,10){\vector(1,0){10}}
\put(24,10){\insertion}
\put(28,10){\line(1,0){20}}
\put(35,10){\vector(1,0){10}}
\put(1,0){$p$}
\put(31,0){$q$}
\thinlines
\put(20,16.7){$T^{xy}$}
\put(8,2){\vector(1,0){10}}
\put(38,2){\vector(1,0){10}}
\end{picture}}\end{picture} \qquad
\frac{i}{4} \left(\gamma^{x} (p^{y}+q^{y})
                + \gamma^{y} (p^{x}+q^{x}) \right)  \,,
\ee
leading to an expression for $\kappa$,
\be
\kappa = -\partial^{2}_{k_{z}}\fsumint_{p}
i^2 (-i)^2\frac{\textrm{Tr}\, \left(
  [(2p{+}k)^x \gamma^y+(2p{+}k)^y \gamma^x]
\nottp
  [(2p{+}k)^x \gamma^y+(2p{+}k)^y \gamma^x]
 [\nottp{+}\nott{k}] \right)}{16 p^2 (p{+}k)^2}
\ee
where prime on the sum-integral indicates that the frequencies
are $(2n+1)\pi T$.  In terms of this fermionic sum-integral,
the equivalent of \Eq{sumint} is
\be
\label{sumintf}
\fsumint_p \frac{(\vec{p}^{\,2})^n}{(p^{2})^{n+1}}
 = \frac{(2n+1)!}{2^{2n} (n!)^2} \: \frac{(-T^2)}{24} \,.
\ee
The rest of the evaluation is straightforward, yielding
\begin{eqnarray}
\kappa= \frac{T^2}{72}\text{\quad for a single flavor}
\end{eqnarray}
Since a Dirac fermion has 4 degrees of freedom, this should be divided
by 4 to get the contribution per degree of freedom.

Next we calculate $\lambda_{3}$ from \Eq{Lambda3}.  The three point
diagram still looks like Figure \ref{scalarlam3diagram1}, but with two
terms depending on whether the
fermion number follows or opposes the indicated momentum flow.
A straightforward evaluation yields a
contribution to $\lambda_3$ of $T^2/24$.

We specialize immediately to the contact term needed in the calculation;
the general expression is not simple.  The formula for $X^{txty}$
is
\be
 X^{txty} = \frac{1}{4} \left(
  e_a^t \frac{\delta T^{ty}}{\delta e_{ax}}
+ e_a^x \frac{\delta T^{ty}}{\delta e_{at}} +
e_a^t \frac{\delta T^{tx}}{\delta e_{ay}} +
e_a^y \frac{\delta T^{tx}}{\delta e_{at}} \right)
\,.
\ee
Using the techniques already introduced, we find after some work that
\be
X^{txty} = \frac{-3}{16} \left(
  \bar\psi \gamma^x \nabla^y \psi
+ \bar\psi \gamma^y \nabla^x \psi
- \nabla^x \bar\psi \gamma^y \psi
- \nabla^y \bar\psi \gamma^x \psi \right)
= \frac{-3}{4} T^{xy} \,.
\ee
The $\langle TX \rangle$ diagram contributing to $\lambda_3$ therefore
gives 3 times the contribution we found for the diagram contributing to
$\kappa$, that is, $T^2/24$.
Adding these two terms, we find
\begin{eqnarray}
\lambda_{3}=\frac{T^2}{12}\text{\quad for a single flavor.}
\end{eqnarray}
Once again, we need to divide by 4 to get the contribution per degree of
freedom.

\subsection{Results}

Since we work, so far, at the free theory level, the result is a
function only of the number of scalar, spinor, and vector degrees of
freedom, which we will write as
$\Nzero$, $\Nhalf$, $\None$.%
\footnote{%
    $\Nzero$ is 1 per real scalar field; $\Nhalf$ is {\sl two} per Weyl
    spinor field, or 4 per Dirac field; and $\None$ is {\sl two}
    per massless spin-1 field (one per spin state).  For
    3-flavor QCD, $\Nzero=0$, $\Nhalf = 4\times 3\times 3 = 36$
    [4 for a Dirac spinor, times three colors times three flavors], and
    $\None = 16$ [2 spin states times 8 color combinations].  For U($N$)
    ${\cal N}{=}4$ SYM theory, $\Nzero = 6N^2$, $\Nhalf=8N^2$,
    and $\None=2N^2$.}
Combining the results of the previous subsections, we find
\bea
\label{eq:kappa}
\kappa & = & \frac{T^2}{288} \left( -4\Nzero + \Nhalf + 8 \None \right)
          +\OO(\sqrt{\alpha}) \,, \\
\label{eq:lambda3}
\lambda_3 & = & \frac{T^2}{48} \left( -4 \Nzero + \Nhalf + 8 \None \right)
          +\OO(\sqrt{\alpha}) \,.
\eea
The other coefficients vanish because the theory is conformal at this
order.  Curiously, at the free level $\lambda_3 = 6 \kappa$ regardless
of the matter content.

We have computed only the leading, coupling-independent
contributions.  We expect the first corrections to $\kappa,\lambda_3$ to
arise at $\OO(\alpha^{\half})$, and the first contributions to the
nonconformal coefficients to be $\OO(\alpha)$.

\section{Lattice implementation}
\label{sec:latt}

Here we will briefly discuss some of the challenges associated with
evaluating the second-order coefficients on the lattice.  One
challenge we foresee is choosing and correctly normalizing the
operators to use on the lattice.  Another challenge is dealing with
(incorrect or divergent) short-distance behavior of the correlators.
We will not discuss the issue of overcoming fluctuations to achieve
good statistics; instead we hope that existing techniques
\cite{Luscher} will prove sufficient.

In general, an operator written in terms of lattice variables will not
correspond to the continuum operator of interest, but will renormalize
and mix with all operators with the same symmetry properties.
For instance, a proposed lattice implementation of $T^{xy}$ will
generically be expressed in terms of the true $T^{xy}$ as
\be
T^{xy}_{\rm latt} = \ZT T^{xy}_{\rm contin} + \sum_n c_n {\cal O}_n^{xy} \,,
\ee
where ${\cal O}_n^{xy}$ are all other operators with the same
symmetries as $T^{xy}$ under the lattice symmetry group, and $\ZT$, $c_n$
are some coefficients.  Generally the operators ${\cal O}_n$ are higher
dimension than $T^{xy}$ and so the $c_n$ will carry positive powers of
the lattice spacing.  Therefore, to the extent that we can take the
continuum limit the ${\cal O}_n$ should be harmless except that they can
introduce short-range contributions to correlators.  However, both the
operation of vacuum subtraction and the small momentum limits associated
with any lattice implementation of
$\partial_{k_z}^2 G(k)|_{k_z \rightarrow 0}$
tend to remove sensitivity to short distance contributions to the
correlators, so we expect this issue to be under control.%
\footnote{%
    The short distance behavior of the stress tensor two-point function
    is $\langle T^{xy}(x) T^{xy}(0)\rangle \sim x^{-8}$.  For a
    dimension-6 operator ${\cal O}_n$, the correlator is
    $\langle T^{xy}(x) {\cal O}_n(0)\rangle \sim a^2 x^{-10}$.  The
    vacuum subtracted value at short distances is ${\cal O}(T^4)$ by OPE
    arguments, see Ref.~\cite{Caron-Huot}; hence
    $\langle T^{xy}(x) {\cal O}_n(0)\rangle_{T} \sim a^2 T^4 x^{-6}$.
    The short distance contribution to
    $\partial_k^2 \langle T{\cal O}_n \rangle(k)$ is
    $\sim \int_x x^2 \langle T(x) {\cal O}_n(0)\rangle
     \sim \int_x a^2 T^4 x^{-4}$ which is ${\cal O}(a^2)$ and at worst
     log UV divergent.  Higher dimension contaminants carry more powers
     of $(a/x)$ and also contribute at order $a^2$.}
The problem
is the renormalization constant $\ZT$, which in general must be determined
nonperturbatively.

To evaluate $\ZT$ it is useful to recall the physical interpretation of
the stress tensor.  If we make a small change to the geometry,
changing $g_{\mu\nu}=\eta_{\mu\nu}$ to
$g_{\mu\nu} = \eta_{\mu\nu} + h_{\mu\nu}$, the action changes from
\be
S_{g=\eta} = \int d^4 x \: {\cal L}_0 \qquad \mbox{to} \qquad
S_{g=\eta+h} = \int d^4 x \left( {\cal L}_0
         - \half h_{\mu\nu} T^{\mu\nu} +{\cal O}(h^2) \right) \,,
\ee
where ${\cal L}_0$ is the Lagrangian density evaluated assuming
$h_{\mu\nu}=0$.
For instance, if we modify the lattice action such that the lattice
spacing in the $x$-direction increases, the change in the action, to
leading order, is $-\half h_{xx}T^{xx}$.
Similarly, if $-h_{xy}T^{xy}$ is added to the action, the geometry becomes
skewed such that the separation between the point $(0,0,0,0)$ and the
point $(0,x,y,0)$ is no longer $\sqrt{x^2+y^2}$ but is
$\sqrt{x^2 + 2xy h_{xy} + y^2}$.

The general strategy for determining the normalization constant $\ZT$ on
the stress tensor is then to include the proposed stress tensor, with small
coefficient $(-c/2)$, in the action, and to see how much it changes the
effective lattice spacing.  For instance, for a diagonal component such
as $T^{xx}$, one can measure correlation lengths along the $x$-axis and
along other lattice axes, or measure the string tension in the $xy$ and
$yz$ planes.  The change in distance determines $h_{xx}$, and the
relation between the proposed $T^{xx}$ and the true one is
$-(c/2) T^{xx}_{\rm proposed} = (-h_{xx}/2) T^{xx}_{\rm true}$ (unless
the change also modifies other axis lengths, in which case the proposed
$T^{xx}$ is a mixture of $T^{xx}$, $T^{yy}$ {\it etc}).

This technique has been well developed for diagonal components of
the stress tensor, see for instance
\cite{Karsch,Caracciolo,Klassen,Karsch2}. To our knowledge it is not
as well developed for the off-diagonal components.  Unfortunately,
all of the Kubo relations we have found, specifically
Eq.~(\ref{kappa}), Eq.~(\ref{lambda3}) and Eq.~(\ref{lambda4}),
involve correlators of off-diagonal components of $T^{\mu\nu}$. But
this is easily fixed by performing rotations in our choice of axes.
For the case of $\kappa$, we make a $\theta=\frac{\pi}{4}$ rotation
in the $(x,y)-$ plane, which transforms \Eq{kappa} to
\be
\label{kappalattice}
\kappa= \frac{1}{4}\lim_{k_{z} \rightarrow 0}
 \frac{\partial^{2}}{\partial k^2_{z}}
  \Big( G_E^{xx,xx}(k) - 2 G_E^{xx,yy}(k) + G_E^{yy,yy}(k) \Big) \,.
\ee
Of course $G_E^{xx,xx}=G_E^{yy,yy}$ at vanishing $k_{x,y}$ by lattice
symmetries, so only one needs to be evaluated.

The correlation function found in \cite{MooreSohrabi} involved all
off-diagonal stress tensors.  Arnold {\it et al} found an expression
involving only $T^{yt}(z)$, $T^{xt}(z)$ \cite{ArnoldVaman}, and we
extend it to the nonconformal case in the Appendix, see
\Eq{lam3xx0y0y}, which we reproduce here:
\be
\label{lambda3_rot1}
\lambda_3 = 2 \kappa^*  -2 \lim_{p^z,q^z\rightarrow 0}
  \partial_{p^z} \partial_{q^z}
   G_{E}^{yy,tx,tx}(p,q) \,.
\ee

\begin{figure}\label{lattice}
\begin{center}
\includegraphics[scale=0.5]{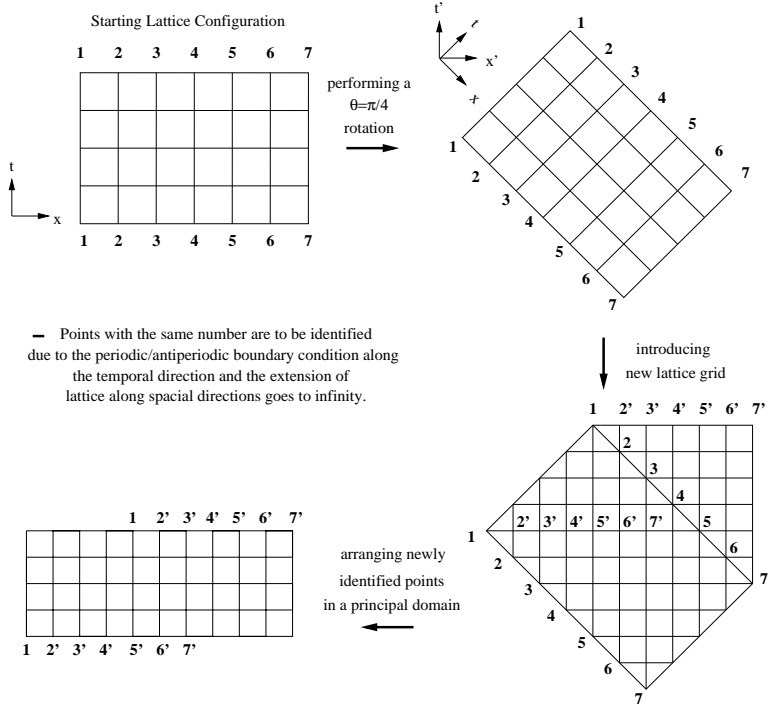}
\end{center}
\caption{How to handle a rotation which mixes a time and a space
  direction, on the lattice.}
\end{figure}

Our expression for $\lambda_3$ still involves the non-diagonal stress tensor $T^{xt}$.
 To re-express it in terms of diagonal terms we must
perform a $45^\circ$ rotation between the $x$ and time axes.
  This requires a change in the implementation of the periodic boundary
conditions in Euclidean space, as illustrated in Figure \ref{lattice}.
As the figure shows, we typically consider a lattice with
principal domain running from $t=0$
to $t=\beta$; equivalently we can say that we consider the field theory
over the whole $x,t$ plane, but with an identification map which equates
every point $(x,t)$ with the point $(x,t+\beta)$.  Introducing rotated
coordinates $x' = (x+t)/\sqrt{2}$ and $t'=(t-x)/\sqrt{2}$,
the identification map equates a point at $(x',t')$ with a
point at $(x'{+}\beta/\sqrt{2},t'{+}\beta/\sqrt{2})$.
We then choose to work on a lattice grid with principal axes along the
$x'$ and
$t'$ directions.  We have to pick a principal domain, that is, a region
of the plane holding exactly one copy of each equivalence class of
points under the periodic identification.  One choice is to stick with
the band of points with $t\in[0,\beta)$; as illustrated in the figure,
the points labeled $1$ are identified, as are the points labeled $2$,
$3$ {\it etc}.  This choice is to consider these points as the boundary
points on the lattice, which are identified with each other.  But the
points labeled $2'$ are also identified, as are the points labeled $3'$
{\it etc}.  Another choice, as illustrated
in the figure, is to choose the band with $t' \in [0,\beta/\sqrt{2})$,
which has these points as the periodically identified boundary.
The identification map relates a point on the bottom edge of this band
with a point on the top edge, {\sl but shifted over} a distance
$\beta/\sqrt{2}$ in the $x'$ direction.  Note that this band is also
narrower than the original band; the inverse temperature $\beta$
corresponds to the space separation of the identified points, not the
extent of the new ``time'' coordinate $t'$.

Therefore, implementing lattice gauge theory on a space where the
periodic identification has a spatial shift corresponds to choosing
axes which lie at an angle with respect to the $(x,t)$ axes.  In
this way we can perform the required $(x,t)$ rotation to make the
stress tensor operators needed in the evaluation of $\lambda_3$
correspond to diagonal components of the stress tensor.
Specifically, in terms of the $x',t'$ coordinates after this final
rotation, $\lambda_3$ is determined by
\be
\label{lattice-con-la3}
\lambda_{3}=2\kappa^{*}-\half\lim_{p_{z},q_{z} \rightarrow 0}\frac{\partial^2}{\partial p_{z}\partial
q_{z}}\left(G_{E}^{tt,tt,yy}(p,q)-2G_{E}^{tt,xx,yy}(p,q) +G_{E}^{xx,xx,yy}(p,q)\right)\,.
\ee
In a completely analogous way, we find that
\bea
\label{lattice-lam4}
\lambda_{4}&\!=\!& \kappa - 2\kappa^{*}- \frac{c^4_{s}}{4}
  \lim_{p,q \rightarrow 0}
  \frac{\partial^2}{\partial p_{x}\partial q_{x}}
 \Big(G_{E}^{tt,tt,xx}(p,q) - G_{E}^{tt,tt,yy}(p,q) \Big) \,.
\eea

\section{Discussion}

Our central results are presented in \Eq{eq:kappa} and \Eq{eq:lambda3}.
The thermodynamic coefficients, unlike the entropy-generating
coefficients $\eta$ {\it etc}, do not diverge in the weak coupling
limit, but remain finite.  Both $\kappa$ and $\lambda_3$ are in
general nonzero.  In particular, the previous observation that
$\lambda_3$ vanishes in strongly-coupled ${\cal N}{=}4$ Super-Yang-Mills
theory \cite{BRSSS,Tata} appears to be an accident.

Curiously, our results for $\kappa$ and $\lambda_3$,
\Eq{eq:kappa} and \Eq{eq:lambda3}, yield zero when we insert the matter
content of ${\cal N}{=}4$ Super-Yang-Mills theory: $\Nzero = 3 \None$
and $\Nhalf = 4 \None$.  Therefore, both
coefficients vanish in the weak-coupling limit.  As we just mentioned,
$\lambda_3$ also vanishes in this theory in the strong coupling limit.
The fact that
$\lambda_3$ vanishes in both limits is suggestive that it is strictly
zero, but this is not the case; it has been shown \cite{SaremiSohrabi}
that $\lambda_3$ is {\sl nonzero} at subleading order in the
large-coupling expansion.  And it is easy to find other examples of
conformal theories where $\lambda_3$ is nonzero.  For instance,
SU($N_c$) gauge theories with fundamental vectorlike matter and with
the number of flavors $N_f$ slightly below $\frac{11}{2}N_c$
have weakly coupled, conformal fixed points \cite{Banks}.
Since such theories are weakly coupled, \Eq{eq:lambda3} applies.  Both
terms are of the same sign, so $\lambda_3$ is certainly not zero for
these conformal gauge theories.

\section*{Acknowledgments}

We thank Simon Caron-Huot and Harvey Meyer for illuminating
discussions.  This work was supported in part by the Natural Sciences
and Engineering Research Council of Canada.

\appendix
\section{Non-conformal hydrodynamics}

\label{appendixA}

In this section we derive Kubo formulae for all the second-order,
thermodynamical transport coefficients of all non-conformal
hydrodynamics. We work in Minkowski space and analytically continue
to Euclidean space. Consider a hydrodynamic system in equilibrium
with some arbitrary background of form
\be
ds^2=\delta_{\mu\nu}dx^{\mu}dx^{\nu}+h_{00}(\vec{x})dt^2
   +h_{0i}(\vec{x})dtdx^{i}+h_{ij}(\vec{x})dx^{i}dx^{j} \,.
\ee
In this curved background the expectation value of the stress
tensor can be expanded about flat space in powers of $h_{\mu\nu}$:
\be
\label{eucl} \langle T^{\mu\nu}_{E}\rangle_{h} = G^{\mu\nu}_{E}
 +\int d^4xG^{\alpha\beta,\mu\nu}_{E}(x,0)\frac{h_{\alpha\beta}(x)}{2}
+\int d^4x d^4y
       G^{\alpha\beta,\gamma\delta,\mu\nu}_{E}(x,y,0)
      \frac{ h_{\alpha\beta}(x)h_{\gamma\delta}(y) }{8}
\ee
where the Euclidean Green functions are defined in \Eq{action_E}.
The corresponding Minkowski space expression is
\be
\label{minko}
\langle T^{\mu\nu}_{r}\rangle_{h}=G^{\mu\nu}_{r}
-\int d^4xG^{\alpha\beta,\mu\nu}_{ar}(x,0)\frac{h_{\alpha\beta}(x)}{2}
 +\int d^4x d^4y G^{\alpha\beta,\gamma\delta,\mu\nu}_{aar}(x,y,0)
  \frac{h_{\alpha\beta}(x)h_{\gamma\delta}(y)}{8} \,,
\ee
where here the retarded Green
functions are the correlation functions of a $T_{r}$ with one or two $T_{a}$,
 as is explained in detail in \cite{MooreSohrabi}. We can
shift from each of these two signatures to the other accordingly by noticing
 that $t$ indices of the Green function are multiplied by a
factor of $i$ and also for each $a$ index we get an extra minus sign.

The following expressions for the curvature tensors will be quite
handy for the derivations of Kubo formulae. To first and second
order of perturbations in the metric of the form,
$g_{\mu\nu}(x)=\eta_{\mu\nu}+h_{\mu\nu}(x)$, we have (borrowing the
result from \cite{Campos})
\begin{eqnarray}
\label{structur}
R_{\alpha\beta\gamma\delta}&=&S_{\sigma[\delta,\beta\gamma]}-h^{\lambda}_{\
\sigma,[\gamma}S_{\lambda\delta],\beta}+
\frac{1}{2}\eta^{\mu\rho}S_{\sigma[\gamma,\mu}S_{\rho\delta],\beta}+
\mathcal{O}(h^3_{\mu\nu})
\\\nonumber
R_{\beta\delta}&=&\frac{1}{2}\eta^{\mu\nu}\left(h_{\mu\delta,\beta\nu}-
h_{\beta\delta,\mu\nu}-h_{\mu\nu,\beta\delta}+h_{\beta\nu,\mu\delta}\right)
-\left(h^{\mu\nu}_{
,\mu}-\frac{1}{2}h^{,\nu}\right)h_{\nu(\delta,\beta)}\\\nonumber&&
+\frac{1}{2}(h^{\mu\nu}h_{\beta\delta,\nu})_{,\mu}+\frac{1}{4}h^{\mu\nu}_{\
,\delta}h_{\mu\nu,\beta}-\frac{1}{4}h^{,\mu}h_{\beta\delta,\mu}
+h_{\mu\delta,\nu}h_{\beta}^{\
[\mu,\nu]}+\frac{1}{2}h^{\mu\nu}h_{\mu\nu,\beta\delta}\\\nonumber&&
-h^{\mu\nu}h_{\mu(\beta,\delta)\nu}+\mathcal{O}(h^3_{\mu\nu})
\\\nonumber
R&=&h^{\alpha\beta}_{,\alpha\beta}-h_{,\alpha}^{\alpha}-
h^{\mu\nu}_{,\mu}h_{\nu\alpha}^{,\alpha}+h^{\mu\nu}_{,\mu}h_{,\nu}+
\frac{3}{4}h^{\mu\nu,\alpha}h_{\mu\nu,\alpha}-\frac{1}{4}h^{,\mu}h_{,\mu}
\\\nonumber&&-\frac{1}{2}h^{\mu\nu,\alpha}h_{\alpha\mu,\nu}-
2h^{\mu\nu}h^{\alpha}_{\mu,\nu\alpha}+h^{\mu\nu}h_{,\mu\nu}+
h^{\mu\nu}{h_{\mu\nu,\alpha}}^{\alpha}+\mathcal{O}(h^3_{\mu\nu})
\end{eqnarray}
and in the above expressions, we have
$S_{\lambda\delta,\beta}=h_{\lambda\delta,\beta}
+h_{\beta\lambda,\delta}-h_{\beta\delta,\lambda}$.

We will write thermodynamic variables $u^\mu,\epsilon,P$ in an
expansion about $h=0$; $u^\mu(x) = \bar u^\mu(x) + u^\mu_{h}(x) +
u^\mu_{h^2}(x)$ and similarly for $\epsilon,P$ (the barred variables
are the $h=0$ values).  It is important to understand the role of
the rest frame; $u^\mu$ need not equal $\bar{u}^{\mu}=(1,0,0,0)$;
rather we must determine $u^\mu_h$ by solving conservation
equations, $\nabla_{\mu}T^{\mu\nu}=0$, consistently and truncating
the expansion. Finally the fluid vector in equilibrium must satisfy
both $\sigma^{\mu\nu}=0$ and $\nabla\cdot u=0$.

We also find it useful to take the trace of the energy-momentum
tensor in the case of non-conformal transport coefficients,
\be
\label{trace}
T^{\mu}_{\mu}=P\left(3-\frac{1}{c_{s}^2}\right)+3\Pi \,.
\ee
To find the
pressure in terms of the background source, we use equations of
motion, in equilibrium we have
\begin{eqnarray}
0&=&\nabla_{\nu}\nabla_{\mu} T^{\mu\nu}\\\nonumber
&=&\nabla_{\nu}\nabla_{\mu}
T_{\text{ideal}}^{\mu\nu}+\nabla_{\nu}\nabla_{\mu}\pi^{\mu\nu}
+\nabla_{\nu}\nabla_{\mu}\left(\Delta^{\mu\nu}\Pi\right)
\end{eqnarray}
where for the ideal fluid, we have
\bea
\label{P}
\nabla_{\nu}\nabla_{\mu}
T_{\text{ideal}}^{\mu\nu}&=&u^{\mu}u^{\nu}\nabla_{\nu}\nabla_{\mu}(\epsilon+P)
+(\epsilon+P)\nabla_{\nu}\nabla_{\mu}(u^{\mu}u^{\nu})\\\nonumber&&
+\nabla_{\mu}(\epsilon+P)\nabla_{\nu}(u^{\mu}u^{\nu})
+\nabla_{\nu}(\epsilon+P)\nabla_{\mu}(u^{\mu}u^{\nu})+\Box P \,.
\eea
Since $u^{i}_{h}, u^{i}_{h^2}\rightarrow 0$ for $\omega\rightarrow
0$, the first term is identically zero, for the second term we get
\begin{eqnarray*}
(\epsilon{+}P)\nabla_{\nu}\nabla_{\mu}(u^{\mu}u^{\nu})
&=&(\epsilon+P)\left(\nabla_{\nu}u^{\nu}\nabla_{\mu}u^{\mu}
  +u^{\nu}\nabla_{\nu}\nabla_{\mu}u^{\mu}
+\nabla_{\nu}u^{\mu}\nabla_{\mu}u^{\nu}
 +u^{\mu}\nabla_{\nu}\nabla_{\mu}u^{\nu}\right)
\\
&\simeq&(\epsilon+P)
  \left((\Gamma_{\nu\lambda}^{\nu}u^{\lambda})^2+
\Gamma_{\nu\alpha}^{\mu}\Gamma_{\mu\beta}^{\nu}u^{\alpha}u^{\beta} +
R_{\sigma\mu}u^{\sigma}u^{\mu}\right)+\mathcal{O}(\omega,h^{3}_{\mu\nu})
\end{eqnarray*}
and we used $[\nabla_{\mu},\nabla_{\nu}]u^{\rho}=R^{\rho}_{\
\sigma\mu\nu}u^{\sigma}$. Similarly, for the third and fourth term
in \Eq{P}, we get
\be
\nabla_{\mu}(\epsilon+P)\nabla_{\nu}(u^{\mu}u^{\nu})
+\nabla_{\nu}(\epsilon+P)\nabla_{\mu}(u^{\mu}u^{\nu})\simeq
2\partial_{\alpha}(\epsilon+P)\Gamma^{\alpha}_{\beta\gamma}
 \bar{u}^{\beta}\bar{u}^{\gamma}+\mathcal{O}(\omega,h^{3}_{\mu\nu})
\,.
\ee
Adding up previous results, finally the pressure reads
\be
\label{pres-ricur}
P=\bar{P} -\frac{\nabla_{\mu}\nabla_{\nu}\pi^{\mu\nu}
+R_{\sigma\mu}\bar{u}^{\sigma}\bar{u}^{\mu}(\epsilon +P+\Pi)+\chi}{\Box}
  -\Pi+\mathcal{O}(\omega,h^{3}_{\mu\nu})
\ee
where $\Box=\sum^3_{i=1}\partial^2_{i}$ and
\be
\chi=(\epsilon+P)
         \left((\Gamma_{\nu\lambda}^{\nu}u^{\lambda})^2+
       \Gamma_{\nu\alpha}^{\mu}\Gamma_{\mu\beta}^{\nu}
       u^{\alpha}u^{\beta} + R_{\sigma\mu}u^{\sigma}u^{\mu}\right)
     +2\partial_{\alpha}(\epsilon+P)\Gamma^{\alpha}_{\beta\gamma}
     \bar{u}^{\beta}\bar{u}^{\gamma}\,.
\ee
This is the generalization of the result in \cite{ArnoldVaman}. As
pointed out in the main text, some Kubo formulae don't
directly relate a transport coefficient to the zero frequency and
momentum limit of Green's functions but they mix these parameters.
Throughout the next section we'll try to find the simplest setup
that can give rise to Green's functions manageable for lattice
calculations.

\subsection{Kubo relation for $\kappa$ and $\xi_{5}$}

We will start the calculation with $\xi_{5}$, by turning on an $h_{xy}(x,y)$
perturbation and evaluating $\langle T^{tt}\rangle$. But this
is $\epsilon$, the energy density of the fluid, we can find it at different
orders of $h_{\mu\nu}$ from \Eq{pres-ricur} or by solving the
equations of motion directly. For illustrative reasons, we do the second approach.
 Since only $h_{xy}(x,y)$ is nonzero we can assume
$u^\mu_h$ is also a function only of $x,y$.  The viscous tensor is
\begin{eqnarray}
\Pi^{xx}&=&\frac{1}{3}\frac{\partial^2h_{xy}(x,y)}{\partial
x\partial
y}\left(\kappa+6\xi_{5}\right)+\frac{2\eta}{3}\left(\frac{\partial
u^{y}_{h}}{\partial y}-2\frac{\partial u^{x}_{h}}{\partial
x}\right)-\zeta\left(\frac{\partial u^{y}_{h}}{\partial
y}+\frac{\partial u_{h}^{x}}{\partial x}\right)\,,\\\nonumber
\Pi^{yy}&=&\frac{1}{3}\frac{\partial^2h_{xy}(x,y)}{\partial
x\partial
y}\left(\kappa+6\xi_{5}\right)+\frac{2\eta}{3}\left(\frac{\partial
u^{x}_{h}}{\partial x}-2\frac{\partial u^{y}_{h}}{\partial
y}\right)-\zeta\left(\frac{\partial u^{y}_{h}}{\partial
y}+\frac{\partial u_{h}^{x}}{\partial x}\right)\,,\\\nonumber
\Pi^{zz}&=&\frac{1}{3}\frac{\partial^2h_{xy}(x,y)}{\partial
x\partial
y}\left(-2\kappa+6\xi_{5}\right)+\frac{2\eta}{3}\left(\frac{\partial
u^{x}_{h}}{\partial x}+\frac{\partial u^{y}_{h}}{\partial
y}\right)-\zeta\left(\frac{\partial u^{y}_{h}}{\partial
y}+\frac{\partial u_{h}^{x}}{\partial x}\right)\,,\\\nonumber
\Pi^{xy}&=&-\eta\left(\frac{\partial u^{y}_{h}}{\partial
x}+\frac{\partial u_{h}^{x}}{\partial y}\right)\,, \quad
\Pi^{xz}=-\eta\frac{\partial u^{z}_{h}}{\partial x}\,, \quad
\Pi^{yz}=-\eta\frac{\partial u^{z}_{h}}{\partial y}
\\\nonumber \Pi&=&-\zeta\left(\frac{\partial u^{y}_{h}}{\partial
y}+\frac{\partial u_{h}^{x}}{\partial
x}\right)+2\xi_{5}\frac{\partial^2h_{xy}(x,y)}{\partial x\partial
y}\,.
\end{eqnarray}
here and through the following sections we neglect dissipative terms
since they will be proportional to the time derivatives of
$h_{\mu\nu}$ or fluid vector in general.

Solving the equations of motion for $\nabla_{\mu}T^{\mu x}=0$,
$\nabla_{\mu}T^{\mu y}=0$, $\nabla_{\mu}T^{\mu z}=0$,
$\nabla_{\mu}T^{\mu t}=0$, we get accordingly,
\begin{eqnarray}
\hspace{-1cm} \frac{\partial P_{h}}{\partial x}
+\frac{\partial u^{x}_{h}}{\partial t}(\bar{\epsilon}+\bar{P})+\frac{\partial^3h_{xy}}{\partial
y\partial x^2}\left(\frac{\kappa}{3}+2\xi_{5}\right)
-\zeta\left(\frac{\partial^2 u^{y}_{h} }{\partial x\partial y}+\frac{\partial^2
u^{x}_{h}}{\partial x^2}\right)
\nonumber \\
-\eta\left(\frac{4}{3}\frac{\partial^2 u^{x}_{h}}{\partial x^2}
+\frac{\partial^2 u^{x}_{h}}{\partial y^2}+\frac{1}{3}\frac{\partial^2
u^{y}_{h}}{\partial x\partial y}\right)&=&0
\nonumber \\
\hspace{-1cm} \frac{\partial P_{h}}{\partial y}+\frac{\partial u^{y}_{h}}{\partial
t}\left(\bar{\epsilon}+\bar{P}\right)+\frac{\partial^3h_{xy}}{\partial x\partial
y^2}\left(\frac{\kappa}{3}+2\xi_{5}\right)
-\zeta\left(\frac{\partial^2 u^{x}_{h} }{\partial x\partial y}
+\frac{\partial^2 u^{y}_{h}}{\partial
y^2}\right)
\nonumber \\
-\eta\left(\frac{4}{3}\frac{\partial^2 u^{y}_{h}}{\partial y^2}
+\frac{\partial^2 u^{y}_{h}}{\partial x^2}+\frac{1}{3}\frac{\partial^2
u^{x}_{h}}{\partial x\partial y}\right)&=&0
\nonumber \\
\frac{\partial u^{z}_{h}}{\partial t}\left(\bar{\epsilon}+\bar{P}\right)
-\eta\left(\frac{\partial^2 u^{z}_{h}}{\partial x^2}+\frac{\partial^2
u^{z}_{h}}{\partial y^2}\right)&=&0
\nonumber \\
\frac{\partial u^{x}_{h}}{\partial x}\left(\bar{\epsilon}+\bar{P}\right)
+\frac{\partial u^{y}_{h}}{\partial
y}\left(\bar{\epsilon}+\bar{P}\right)+\frac{\partial \epsilon_{h}}{\partial t}&=&0 \,.
\end{eqnarray}
Since we are interested in the zero frequency limit all time
derivatives are zero. As we can see in the above equations terms
proportional to the metric perturbation appear, which act as a source
for $P$ and for $u^\mu$. Higher order terms that include the
interaction of the fluid vector with background perturbation have been
neglected. From the first two equations we obtain
\be
P_{h}=-\left(\frac{\kappa}{3}+2\xi_{5}\right)
 \frac{\partial^2h_{xy}}{\partial y\partial x}+\mathcal{O}(\partial_{t})
\ee
and we know that pressure and energy density are related through
$P=c^2_{s}\epsilon$. Similarly we get
\be
u^{x}_{h}=\frac{1}{3}\frac{\omega h_{xy} q^2_{x}q_{y}(6\xi_{5}+\kappa)}
     {c_{s}^2(q^2_{x}+q_{y}^2)}(\bar{\epsilon}+\bar{P})
     +\mathcal{O}(\omega^2)\,,\quad
u^{y}_{h}=\frac{1}{3}\frac{\omega h_{xy} q^2_{y}q_{x}(6\xi_{5}+\kappa)}{c_{s}^2(q^2_{x}+q_{y}^2)}
   (\bar{\epsilon}+\bar{P})+\mathcal{O}(\omega^2)\,.
\ee
Finally, from \Eq{minko}, we have
\be
\frac{-1}{c_{s}^2}\left(\frac{\kappa}{3}+2\xi_{5}\right)
  \frac{\partial^2h_{xy}}{\partial y\partial x}
=-\frac{1}{2}\int d^4xG^{\alpha\beta,tt}_{ar}(x,0)h_{\alpha\beta}(x)\,.
\ee
Fourier transforming and taking the variation of both sides
with respect to $h_{xy}$, we get
\be
\xi_{5}=-\frac{c_{s}^2}{2}\lim_{k_{x},k_{y} \rightarrow 0}
  \frac{\partial^2}{\partial k_{x}\partial k_{y}}
  G_{ar}^{xy,tt}(k)-\frac{\kappa}{6}
\ee
and we know that
\be
\lim_{k_{0} \rightarrow 0}G_{ar}^{xy,tt}(k_{0},\mathbf{k})
 =+G_{E}^{xy,tt}(k_{0}=0,\mathbf{k})
\ee
So finally in Euclidean space we have
\be
\label{new-iden}
\xi_{5}=-\frac{c_{s}^2}{2}\lim_{k_{x},k_{y} \rightarrow 0}
\frac{\partial^2}{\partial k_{x}\partial k_{y}}
  G_{E}^{xy,tt}(k)-\frac{\kappa}{6}
\,.
\ee

Similarly the Kubo relation for $\kappa$ in terms of off-diagonal
components of stress-tensors is given by
$\kappa=-\partial^{2}_{k_{z}}G_{ar}^{xy,xy}(k)$.
In terms of the
Euclidean correlator this is
$\kappa=\partial^{2}_{k_{z}}G_{E}^{xy,xy}(k)$.

\subsection{Kubo relation for $\kappa^{*}$ and
$\xi_{6}$}

To find a Kubo relation for $\xi_{6}$, we use the perturbation $h_{tt}(z)$.
This choice, shifts the local rest frame by $u^{t}=1+1/2
h_{tt}+\mathcal{O}(h^2)$. Expanding the fluid vector in terms of metric
perturbation $h_{\mu\nu}$, we find the following viscous tensors,
\begin{eqnarray}
\Pi^{xx}&=&\Pi^{yy}=\frac{1}{6}\frac{\partial^2 h_{tt}}{\partial z^2}
\left(-2\kappa^{*}+\kappa-3\xi_{6}+6\xi_{5}\right)
\nonumber \\
\Pi^{zz}&=&\frac{1}{6}\frac{\partial^2 h_{tt}}{\partial z^2}
\left(4\kappa^{*}-2\kappa-3\xi_{6}+6\xi_{5}\right)
\nonumber \\
\Pi&=&-\frac{\xi_{6}}{2}\frac{\partial^2 h_{tt}}{\partial z^2}
+\xi_{5}\frac{\partial^2 h_{tt}}{\partial z^2}\,.
\end{eqnarray}
If we assume that the hydrodynamic waves are only functions of $z$
then for $\nabla_{\mu}T^{\mu z}=0$ we have
\be
 0=\frac{\partial P_{h}}{\partial
z}+\frac{\partial u^{z}_{h}}{\partial t}(\bar{\epsilon}+\bar{P})
-\frac{1}{2}\frac{\partial h_{tt}}{\partial
z}(\bar{\epsilon}+\bar{P})+\frac{\partial^3 h_{tt}}{\partial z^3}
\left(-\frac{\xi_{6}}{2}+\frac{2\kappa^{*}}{3}
 -\frac{\kappa}{3}+\xi_{5}\right)
\ee
and once again the last term acts as a source for pressure.
For $\langle T^{tt}\rangle$ we have
\be
\label{T00}
\langle T^{tt}\rangle_{h} = \frac{\left(\bar{\epsilon}+\bar{P}\right)}
  {2 c_{s}^2}h_{tt}(z) +\left(\frac{\xi_{6}}{2}
  -\frac{2\kappa^{*}}{3}+\frac{\kappa}{3}-\xi_{5}\right)
  \frac{\partial^2_{z}h_{tt}(z)}{c_{s}^2} \,.
\ee
The first term in the above relation is a pure gauge. For the second
term from linear-response we have
\be
\left(\frac{\xi_{6}}{2}-\frac{2\kappa^{*}}{3}
   +\frac{\kappa}{3}-\xi_{5}\right)
  \frac{\partial^2_{z}h_{tt}(z)}{c_{s}^2}
 = -\frac{1}{2}\int d^4xG^{\alpha\beta,tt}_{ar}(x,0)h_{\alpha\beta}(x) \,.
\ee
 The corresponding Kubo formula for $\xi_{6}$ will
be
\be
\xi_{6}=2\xi_{5}+\frac{4\kappa^{*}}{3}-\frac{2\kappa}{3}
 +\frac{c^2_{s}}{2}\lim_{k_{z} \rightarrow 0}
  \frac{\partial^2}{\partial k_{z}^2}G^{tt,tt}_{ar}(k)
\ee
and accordingly in Euclidean space using
$\lim_{k_{t} \rightarrow 0}G_{ar}^{tt,tt}(k_{t},\mathbf{k})
 =-G_{E}^{tt,tt}(k_{t}=0,\mathbf{k})$, it can be rewritten as
\be
\xi_{6}=2\xi_{5}+\frac{4\kappa^{*}}{3}-\frac{2\kappa}{3}
 -\frac{c^2_{s}}{2}\lim_{k_{z} \rightarrow 0}
  \frac{\partial^2}{\partial k_{z}^2}G^{tt,tt}_{E}(k) \,.
\ee
This formula determines a linear combination of $\xi_6$ and
$\kappa^*$. To get both coefficients separately we need to look for
another relation for $\kappa^*$.  We do so by investigating an
off-diagonal component of energy-momentum tensor. We have
\be
\langle T^{xy}\rangle=(\epsilon+P)u^{x}u^{y}
  +P g^{xy}+\Pi^{xy} \,.
\ee
Since $\kappa^*$ is a coefficient involving the curvature tensor,
which first arises at linear order in $h$, we need only work to this
order, in which case the first term is zero; and if we choose $h_{xy}$
nonzero then $P g^{xy}$ and $\Delta^{xy} \Pi$ are also zero.
Therefore we consider $h_{tt}(x,y)$. Then the only contribution comes
from $\pi^{xy}$, after expansion in the orders of $h_{tt}$ and
$u^{\mu}(x)=\bar{u}^{\mu}(x)+u_{h}^{\mu}(x)+\mathcal{O}(h^2)$, we find
\be
\pi^{xy}=\frac{\partial^2 h_{tt}}{\partial x\partial y}
  \left(\kappa^{*}-\frac{\kappa}{2}\right)
\ee
If we use \Eq{minko} and Fourier transforming, we can write the Kubo
formula in Euclidean space as
\be
\kappa^{*}=\frac{\kappa}{2}+\frac{1}{2}
  \lim_{k_{x},k_{y} \rightarrow 0}\frac{\partial^2}
   {\partial k_{x}\partial k_{y}}G_{E}^{xy,tt}(k)\,.
\label{kapstar}
\ee

\Eq{kapstar} and \Eq{new-iden} involve the same Green function, so we
find a relation between $\xi_5$ and $\kappa^*$, specifically
\be
\xi_{5}=-c^2_{s}\kappa^{*}
  +\frac{\kappa}{2}\left(c^2_{s}-\frac{1}{3}\right)\,.
\ee

\subsection{Kubo relation for $\lambda_{3}$ and $\xi_{3}$}
\label{append-xi3}

Now we turn to nonlinear transport coefficients, where we must work to
second order in $h$. We begin with $\lambda_{3}$, which is the traceless
contribution arising at second order in vorticity.  Vorticity is generated
 by a nonvanishing value of $h_{ti,j}$; specifically the vorticity
term for which $\lambda_3$ is a parameter (see \Eq{T_order2}) is
\be
\Omega^{\langle i}_{\ \lambda}\Omega^{j\rangle\lambda}=\frac{1}{12}
\left(\delta^{ij}\delta_{mn}-3\delta^{i}_{m}\delta^{j}_{n}\right)
 \epsilon^{mkl}\epsilon ^{nrs}\partial_{k}h_{lt}\partial_{r}h_{st} \,.
\ee
The easiest way to proceed \cite{MooreSohrabi,ArnoldVaman} is to
consider an off-diagonal component of $T$, such as $T^{xy}$; then
complications involving the pressure (such as those of the last
subsections) do not arise.  However as we have discussed it is most
convenient on the lattice to use a relation involving a diagonal
component of the stress tensor.  Therefore we will instead consider the
vorticity-related contributions to $T^{xx}$:
\be
\label{dummyT} \langle T^{xx}\rangle = (\epsilon+p)u^{x}u^{x} +pg^{xx}+\Pi^{xx} \,.
\ee
Since we have to keep all terms to the order of $\mathcal{O}(h^2)$,
we need to know $u^{x}_{h}$, $P_{h^2}$, and $\Pi^{xx}_{h^2}$.  We will
consider nonvanishing $h_{ty}(z)$, which is general enough for $\Pi^{xx}$
to contain a $\lambda_3$ dependent term.  For this choice $u^x_h$
vanishes.  To find the contribution of $P_{h^2}$, we look into \Eq{pres-ricur}.
 Since $\Gamma^{\nu}_{tt}$ is zero for our metric
perturbation, the only possible contributions come from
$\partial_{\mu}\partial_{\nu}\pi^{\mu\nu}$ and $\Pi$. All derivatives other than
$\partial_{z}$ are zero, so finally we have
\be
\label{ppp} P_{h^2}=-\pi^{zz}-\Pi=-\Pi^{zz} \,.
\ee
Calculating $\Pi^{zz}$, $\Pi^{xx}$ and
recalling that $u^{\mu}_{h}=0$, the final result reads
\begin{eqnarray*}
\Pi^{zz}&=&-h_{ty}\frac{\partial^2 h_{ty}}{\partial z^2}
 \left(2\xi_{5}+\frac{\kappa}{3}\right)
 +\left(\frac{\partial h_{0y}}{\partial z}\right)^2
 \left(-\frac{3\xi_{5}}{2}+\frac{\lambda_{3}}{12}+\frac{\xi_{6}}{2}
  -\frac{\kappa^{*}}{6}+\frac{\xi_{3}}{2}\right)+\mathcal{O}(h^3)
  \\
\Pi^{xx}&=&2h_{ty}\frac{\partial^2 h_{ty}}{\partial z^2}
  \left(-\xi_{5}+\frac{\kappa}{3}\right)
  +\left(\frac{\partial h_{ty}}{\partial z}\right)^2
  \left(-\frac{3\xi_{5}}{2}-\frac{\lambda_{3}}{6}+\frac{\xi_{6}}{2}
  +\frac{\kappa^{*}}{3}+\frac{\xi_{3}}{2}\right)+\mathcal{O}(h^3) \,.
\end{eqnarray*}
Now for \Eq{dummyT} we can write
$\langle T^{xx}\rangle=P+\Pi^{xx}=\bar P-\Pi^{zz}+\Pi^{xx}$, which reads
\be
\langle T^{xx}\rangle - \bar
P = h_{ty}\frac{\partial^2 h_{ty}}{\partial z^2}\kappa+\left(\frac{\partial h_{ty}}{\partial
z}\right)^2\left(-\frac{\lambda_{3}}{4}+\frac{\kappa^{*}}{2}\right)
 +\mathcal{O}(h^3) \,.
\ee
For the response of the three-point Green's function, we have
\be
h_{ty}\frac{\partial^2 h_{ty}}{\partial z^2}\kappa
  +\left(\frac{\partial h_{ty}}{\partial z}\right)^2
  \left(-\frac{\lambda_{3}}{4}+\frac{\kappa^{*}}{2}\right)
  =\frac{1}{8}\int d^4x d^4y
   G^{\alpha\beta,\gamma\delta,xx}_{aar}(x,y,0)
   h_{\alpha\beta}(x)h_{\gamma\delta}(y)
\ee
which results in the following Kubo formula for $\lambda_{3}$:
\be
\lambda_{3}=2\kappa^{*}+2\lim_{p_{z},q_{z} \rightarrow 0}
  \frac{\partial^2}{\partial p_{z}\partial q_{z}}
  G_{aar}^{ty,ty,xx}(p,q)\,,
\ee
and in Euclidean space through
$\lim_{p_{0},q_{0} \rightarrow 0}G_{aar}^{ty,ty,xx}(p,q)
=-G_{E}^{ty,ty,xx}(\mathbf{p},\mathbf{q})$, we get
\be
\lambda_{3}=2\kappa^{*}-2\lim_{p_{z},q_{z} \rightarrow 0}
   \frac{\partial^2}{\partial p_{z}\partial q_{z}}
   G_{E}^{ty,ty,xx}(p,q) \,.
\label{lam3xx0y0y}
\ee
In the main text we use this expression but with $x\leftrightarrow y$.

It's quite easy to find a Kubo relation for $\xi_{3}$ now, since we have a
 relation for $\lambda_{3}$. From the trace of stress-tensor we have
\be
T^{\mu}_{\mu}=P\left(3-\frac{1}{c_{s}^2}\right)+3\Pi\,.
\ee
Using \Eq{ppp} and the expressions for $\Pi^{zz}$ and $\Pi$ that can be
calculated as
\be
 \Pi=-2h_{ty}\frac{\partial^2 h_{ty}}{\partial z^2}\xi_{5}
  +\frac{1}{2}\left(\frac{\partial h_{ty}}{\partial z}\right)^2
  \left(\xi_{6}-3\xi_{5}+\xi_{3}\right)+\mathcal{O}(h^3)\,,
\ee
and from the response of the retarded Green's function and then analytically
 continuing to Euclidean space, we finally obtain the Kubo
relation for $\xi_{3}$:
\be
\xi_{3}=3\xi_{5}-\xi_{6} +\left(c_{s}^2-\frac{1}{3}\right)
  \left(\frac{\lambda_{3}}{2}-\kappa^{*}\right)
+c^2_{s}\lim_{p_{z},q_{z} \rightarrow 0}
 \frac{\partial^2}{\partial p_{z}\partial q_{z}}
 G_{E}^{ty,ty,\mu\mu}(p,q) \,.
\ee
\subsection{Kubo relation for $\lambda_{4}$ and
$\xi_{4}$}\label{append-xi4}

The coefficients $\lambda_4$ and $\xi_4$ arise when the entropy,
and therefore temperature, vary in space.  For this to occur in
equilibrium, the gravitational potential $h_{tt}$ must vary in space.
  Therefore we consider perturbations of $h_{tt}(x,y)$ to second
order. The off-diagonal component of the stress-tensor, $\langle T^{xy}\rangle$, is
\be
\pi^{xy}=\pi^{xy}_{h}-\frac{\kappa}{2}
h_{tt}\frac{\partial^2h_{tt}}{\partial x\partial y} +\kappa^{*}h_{tt}
\frac{\partial h_{tt}}{\partial x\partial
y}+\left(\frac{\lambda_{4}}{4c^4_{s}}-\frac{\kappa}{4} +\frac{\kappa^{*}}{2}\right)
\frac{\partial h_{tt}}{\partial x}\frac{\partial
h_{tt}}{\partial y}
\ee
 that after analytic continuation reduces to
\be
\lambda_{4}=-\frac{c^{4}_{s}}{2}\lim_{p_{x}, q_{y} \rightarrow 0}\frac{\partial^2}{\partial p_{x}\partial
q_{y}}G_{E}^{tt,tt,xy}(p,q)-2\kappa^{*}+\kappa\,.
\ee

To find $\xi_4$ we evaluate $\langle T^{\mu}_{\mu}\rangle$ for a perturbation of $h_{tt}(z)$.
>From \Eq{trace}, we find the pressure through
\Eq{pres-ricur}. We have $\Gamma^{t}_{tz}=-1/2\partial_{z}h_{tt}$,
 $\Gamma^{z}_{tt}=-1/2\partial_{z}h_{}$, and
$R_{tt}=-1/2\partial^{2}_{z}h_{tt}-1/4(\partial_{z}h_{tt})^2$, which result in
\be
(\epsilon{}+P)=(\bar{\epsilon}{}+\bar{P})+\left(1+\frac{1}{c^2_{s}}\right)
 \left(\frac{\bar{\epsilon}
+\bar{P}}{2}h_{tt}+\left(\frac{\xi_{6}}{2} -\frac{2\kappa^{*}}{3}+\frac{\kappa}{3} -\xi_{5}\right)\frac{\partial^2 h_{tt}}
    {\partial z^2}\right)\,,
\ee
from \Eq{T00}, and
\bea
\label{pi}
(\epsilon+P)R_{\sigma\mu}\bar{u}^{\sigma}\bar{u}^{\mu}
&=&-\frac{(\bar{\epsilon}+\bar{P})}{4}\left(\frac{\partial h_{tt}} {\partial z}\right)^2
-\frac{(\epsilon+P)}{2} \frac{\partial^2
h_{tt}}{\partial z^2} \,,
\\ \nonumber
R_{\sigma\mu}\bar{u}^{\sigma}\bar{u}^{\mu}\Pi& = & -\frac{1}{2}\frac{\partial^2 h_{tt}}{\partial z^2}
 \left(\xi_{5}-\frac{\xi_{6}}{2}\right)\frac{\partial^2 h_{tt}}
   {\partial z^2} \,,
\\ \nonumber
\chi & = & \chi_{h} + \frac{\bar{\epsilon}+\bar{P}}{4}
  \left(\frac{\partial h_{tt}}{\partial z}\right)^2
 -\frac{\partial (\epsilon+P)_{h}}{\partial z}
  \frac{\partial h_{tt}}{\partial z}
\\ \nonumber
\Pi^{zz}&=&\Pi^{zz}_{h}+\left(\frac{\xi_{4}}{4c_{s}^4}-\frac{\xi_{6}}{4}
+\frac{3\xi_{5}{+}2\kappa^*{-}\kappa}{6}+\frac{\lambda_{4}}{6c^{4}_{s}} \right)
 \left(\frac{\partial h_{tt}}{\partial z}\right)^2
 \\ \nonumber &&
 +\left(\xi_{5}{-}\frac{\xi_{6}}{2}+\frac{2\kappa^{*}-\kappa}{3}\right) h_{tt}
 \frac{\partial^2 h_{tt}}{\partial z^2}\,.
\eea
After Fourier transforming and straightforward simplifications, the pressure reads
\bea
P_{h^2}&=&\frac{\bar{\epsilon}+\bar{P}}{4}
+\frac{\bar{\epsilon}+\bar{P}}{8}\left(1+\frac{1}{c^2_{s}}\right)
+\frac{p_{z}q_{z}}{4}\left(\frac{2\kappa^{*}}{3}+\frac{2\lambda_{4}}{3c^{4}_{s}}
 -\frac{\xi_{6}}{2}-\frac{\kappa}{3}+\xi_{5}+\frac{\xi_{4}}{c^4_{s}}\right)
\nonumber \\
&& -\frac{p_{z}q_{z}}{4}\left(1+\frac{1}{c^2_{s}}\right)\left(\frac{\xi_{6}}{2}
-\frac{2\kappa^{*}}{3}+\frac{\kappa}{3}-\xi_{5}\right)
 +\mathcal{O}(p^2,q^2) \,.
\eea
Inserting the above expressions for $P$ and $\Pi$ from \Eq{pi}, finally we get
\bea
\xi_{4} & = & -\frac{c_{s}^6}{2}\lim_{p_{z}, q_{z}
\rightarrow 0}
  \frac{\partial^2}{\partial p_{z}\partial q_{z}}
  G_{E}^{tt,tt,\mu\mu}(p,q)
 +6c^6_{s}\left(\frac{\xi_{6}}{2} -\xi_{5}\right)
\\ \nonumber & &
 +4c^6_{s}\left(3-\frac{1}{c^2_{s}}\right)
 \left(+\frac{1}{4}\left(\frac{2\kappa^{*}}{3}+\frac{2\lambda_{4}}{3c^{4}_{s}}
 -\frac{\xi_{6}}{2}-\frac{\kappa}{3}+\xi_{5}+\frac{\xi_{4}}{c^4_{s}}\right)
 \right. \\ \nonumber &&\left.
 -\frac{1}{4}\left(1+\frac{1}{c^2_{s}}\right)\left(\frac{\xi_{6}}{2}-\frac{2\kappa^{*}}{3}+
\frac{\kappa}{3}-\xi_{5}\right)\right) \,.
\eea

\end{document}